\documentclass[aps,prc,reprint,amssymb,amsmath,superscriptaddress, floatfix,10pt]{revtex4-1}
\usepackage{mathptmx}
\usepackage{graphicx}
\usepackage{color}  
\makeindex
\newcommand{\textblue}{\textcolor[rgb]{0.00,0.07,1.00}}

\begin{document}

\title{Unified description of superconductivity in neutron stars}
\author{Dmitry Kobyakov}
\email{dmitry.kobyakov@appl.sci-nnov.ru}
\affiliation{Institute of Applied Physics of the Russian Academy of Sciences, 603950 Nizhny Novgorod, Russia}
\date{August 17, 2023}

\begin{abstract}
In this paper, I study the location and symmetry of superconducting protons.
Solving the Tolman-Oppenheimer-Volkoff (TOV) equations based on the unified Barcelona-Catania-Paris-Madrid equation of state (BCPM EoS) and on the pairing gap calculations by Lim and Holt \cite{LimHolt2021}, I find that roughly 500 meters of the liquid core (with isotropic and continuous  symmetry) and roughly 100-150 meters of the core-crust interface (with anisotropic symmetry) are superconducting, while the rest of the star is normal.
To specify whether the superconducting symmetry is discreet in the pasta phase, I study the coexistence of the saturated nuclear and the pure neutron matter using EoS based on the chiral effective field theory (ChEFT).
I find that the maximum pressure at coexistence is $P_{*}\simeq0.5\;{\rm MeV\,fm^{-3}}$.
To verify the precision of the coexistence calculations I evaluate the surface and the Coulomb corrections using the compressible liquid drop model.
I calculate the proton tunneling rate in the perfectly ordered slab region of the pasta phase and conclude that for the chosen EoS, the proton supercurrent tunneling between the adjacent slabs is negligible and the slab region should be described as a discreet symmetry system of quasi two-dimensional layers.
\end{abstract}

\maketitle

\section{Introduction}
Theoretical explanation of a wide range of observable phenomena in neutron stars relies on assumptions about the electrical conductivity of its matter.
Rich structure of neutron stars includes a completely ionized crystal lattice, superfluid liquid crystals in the crust, three-component liquid plasma in the outer core and yet unknown plasma in the inner core.
This richness induces complicated properties of the electrical conductivity throughout the entire star and considerably complicates the stellar magnetism.

An additional complication is provided by superconductivity of protons.
In 1969, the existence of proton superconductivity in neutron stars has been considered by Baym, Pethick and Pines \cite{BPP1969}.
Since then, the standard picture has been that superconducting protons uniformly fill the stellar core and symmetry of the order parameter is isotropic and continuous.
However in 2018, Kobyakov \cite{Kobyakov2018} and Kobyakov and Pethick \cite{KobyakovPethick2018} have shown that the effective symmetry of superconducting protons at the core-crust boundary might be anisotropic and discreet.
Based on the assumption of the discreet superconducting symmetry, which needed confirmation from microscopic physics, Kobyakov \cite{Kobyakov2018} has shown that the slab region is interesting for astrophysical applications due to a possible pressure drop resulting when the superfluid neutron entrainment with protons is taken into account.
In 2021, Lim and Holt \cite{LimHolt2021} have shown that the superconductor gap energy closes at higher densities.
Thus, it is necessary to specify the properties of superconductivity in neutron stars.
Its role has been studied in numerous works; I briefly review some of them.
Influence of superconductivity on the structure of the stellar magnetic field was studied in \cite{Lander2013,HenrikssonWasserman2014}.
The role of superconductivity in evolution of the magnetic field was studied in \cite{MuslimovTsygan1985,GraberEtAl2015,PassamontiEtAl2017}.
Relations of magnetism and the neutrino cooling processes were studied in \cite{SinhaSedrakian2015}.
The role of superconductivity on the pulsar glitches was explored in \cite{HaskellEtAl2013}.
Superconductivity in the context of quasi-periodic oscillations (QPOs) in the afterglows of giant magnetar flares was addressed in \cite{Levin2006,Bretz2021}.

The importance of observational probe offered by the phenomenon of QPOs is remarkable because the observable frequency spectrum of the QPOs contains useful signatures of the internal magnetar mechanisms and structure within deep layers of neutron star \cite{Levin2006,Bretz2021,SotaniEtAl2017,Elenbaas2016}.
Theory of QPO is also expected to help constrain structure and dynamics of the inner core of neutron stars, where the matter energy density $\rho$ is more than about $1.1\rho_0$, with $\rho_0=m_nn_0$, $m_n$ is the mass of neutron and $n_0=0.16\;{\rm fm}^{-3}$ is the baryon number density at nuclear saturation.
A fundamental difficulty is that EoS of the inner core cannot be determined from the information solely about the finite nuclei.

The effective degrees of freedom in the inner core are likely to be not the effective ones associated with the ordinary nucleon matter.
As a result, the existing theoretical models provide mutually incoherent predictions.
Fortunately, investigation of the inner core is not hopeless due to a growing body of observations of the astrophysical mechanisms which presumably probe the inner core.
Insisting on the coherence of various models on the global scale of the star requires unification of the description of the stellar structure.

The global stellar structure is intimately related to the magnetism, which is a standard base of observable phenomena.
As the magnetism and superconductivity are also intimately related, the description of magnetism requires the details of superconductivity throughout the entire star.
Thus a problem of unified description of superconductivity emerges.
To address this question, in this paper I investigate the problem of the exact location of the superconducting regions, their symmetry and astrophysical consequences.

The first step is to calculate the structure of neutron star matter in the gravitational field, which will be done by solution of TOV equations with a unified EoS of dense matter.
Equation of state valid throughout the star have been addressed in the literature, for instance, by Douchin and Haensel \cite{DouchinHaensel2001} and by Sharma, Centelles, Vinas, Baldo and Burgio \cite{SharmaEtAl2015}.
The work \cite{SharmaEtAl2015} provides the EoS widely known as the BCPM EoS.
For aims of this paper I will also use other EoS: the parametric EoS proposed by Baym, Bethe and Pethick \cite{BBP1971} and its more recent version proposed by Hebeler, Lattimer, Pethick and Schwenk based on the ChEFT \cite{HebelerEtAl2013}.
These are applicable at arbitrary isospin asymmetry out of beta equilibrium and will be used together with the polytropic extension into the inner core, which will allow to place the predictions of the BCPM EoS along with the predictions of soft, intermediate and stiff EoS based on ChEFT with the polytropic extension.

Following this first step, I will combine the structure calculations with the available calculations of the proton pairing gap energy as function of the proton Fermi wavenumber after Lim and Holt \cite{LimHolt2021}.
In this way it will become possible to find the exact location of the superconducting regions.

As a next step, I will investigate possible types of symmetry of the superconductor.
For this purpose, a more detailed approach is required.
In particular, new symmetries are expected \cite{Kobyakov2018}, \cite{KobyakovPethick2018} in the pasta phases due to the periodicity of the structure, if the structure is ordered.
At present, whether the order is present is not known \cite{PethickZhang2022}.
Below I will explore the perfectly ordered state and outline a strategy to advance understanding of ordering in the pasta structure.

Serious difficulties for the theory arise because the modern picture of the pasta phases is strongly model dependent, and the predictions are mutually incoherent.
Origins of the incoherence in the pasta predictions arise due to various factors, and the main factors seem to be the existence of a great number of EoS for the range of baryon densities corresponding to the crust-core transition and the fact that the different pasta configurations differ by less than one per cent from the bulk nuclear energy involved in the problem.
Thus the effect of the pasta phase is quite subtle, and the numerical calculations predicting the pasta structure should explicitly analyze and display the involved numerical errors.
Reduction of the incoherence is an actual problem, which will be addressed elsewhere.
A promising approach is the compressible liquid drop model, which can be used with various EoS and thus, the predictions may be systematically ordered.
The basic idea of this approach is to consider the coexistence of the pure neutron matter and the dense (saturated) nuclear matter \cite{BBP1971} while taking into account the surface and Coulomb corrections \cite{RavenhallEtAl1983,HashimotoEtAl1984,WatanabeEtAl2000,Vinas1998,Vinas2017}.

In this paper I will consider only the slab type of pasta phases, however, other lattices are possible, such as the rod-like nuclei, rod-like bubbles and the spherical bubbles.
These phases, if the pasta is ordered, seem to be less interesting than the slab-like region in the context of the force  considered in \cite{Kobyakov2018} and associated with the simultaneous presence of the superfluid-superconducting velocity lag and the magnetic field.
The reason is the corresponding structure of the proton supercurrent in those phases.
From this perspective, the bubble phase is not very different from the uniform phase, because the supercurrents are free to flow in any of the three Cartesian directions.
As to the rod-like nuclei, the supercurrent structure is lower dimensional like in the slab region, however, the rod-like nuclei contain considerably less baryons in the dense phase.
As a result, the force considered in \cite{Kobyakov2018} if applies to the rod-like nuclei does not induce as much stress in the crust as compared with the slab-like nuclei, where the stress might be capable to shatter the crust.
However, if the slabs are not energetically favorable and are not realized then the rod-like nuclei might provide the effect discussed in \cite{Kobyakov2018} in a reduced form.

Ordering of the slab region is not the only necessary condition required by the peculiar force found in the pasta phases \cite{Kobyakov2018}.
As explained in \cite{Kobyakov2018}, for the effect to be viable, the slabs must be quasi two-dimensional in the sense that the proton flow between the adjacent slabs should be negligibly small.
In this paper I will advance understanding of this physical picture by calculating the amplitude of proton tunneling between the slabs as function of pressure.
The calculation requires to find the proton chemical potential in the pure neutron matter, which together with the proton chemical potential in the dense phase at coexistence will allow to evaluate the potential energy barrier experienced by the protons between the slabs.

Since the pressure range within the slab region is not known exactly, I will calculate the barrier height for a relevant range of pressures.
It should be noted that the pressure in the BCPM EoS is computed with only the spherical nuclei.
The differences with other more favoured configurations in the bottom of the inner crust are very small, 1-2 keV or less.
Strictly speaking the BCPM EOS in the bottom of the inner crust does not contain contributions from the planar or cylindrical configurations
although BCPM predicts slabs in the short range of densities between 0.076 and 0.082 ${\rm fm}^{-3}$ as it can be seen in table 5 of \cite{SharmaEtAl2015}.

Finally, I will discuss the uncertainties related to the slab region of the pasta phases.
Generally, the uncertainties include
\begin{itemize}
  \item (i) the radial position and existence of the layer with slabs,
  \item (ii) order or disorder of the slabs,
  \item (iii) the thickness and separation of the superconducting slabs,
  \item (iv) orientation of the structure with respect to the magnetic field.
\end{itemize}

In this paper I will systematically investigate the point (iv) and discuss the magnetic properties of the slab region at different angles between the magnetic field and the slab structure.
Other points will be addressed elsewhere.

The structure of this paper is the following.
In Section II, I describe the input data needed to calculate the spatial profile of superconducting matter and details of the coexistence.
Section III is reserved for the numerical results.
The magnetism of superconducting neutron stars is discussed in Section IV.
Conclusions are placed in Section V.

\section{Spatial profile of superconducting matter}
The location of the superconducting matter in neutron stars is determined by a combination of the spatial profile of the matter mass density $\rho(r)$ and the dependence of the superconducting transition temperature (or, equivalently $\Delta_p$) on $\rho$.
I will work with a static neutron star with spherical symmetry and thus the only spatial coordinate is the stellar radius $r$.
Influence of the superconductivity on the pressure is negligible, therefore one may solve the problem in two steps.
The first step is to find $\rho(r)$ and the second step, assuming constant temperature across the region containing the superconducting matter, is to translate the dependence of $\Delta_p(\rho)$ into the dependence $\Delta_p(r)$.
{This approach is analogous to the local density approximation.
Similarly, it is possible to find the neutron superfluidity layer of a neutron star, if the spatial dependence of the neutron gap energy were used.}

The function $\rho(r)$ is determined by a balance of the gravitational force and the repulsive force due to pressure $P(r)$ of the stellar matter, which is encoded in the TOV equations:
\begin{eqnarray}
\label{TOVeq1}&& \frac{dP}{dr}=G\frac{mc^2+4\pi r^3P}{2Grmc^2-c^4r^2}\left(\rho c^2+P\right),\\
\label{TOVeq2}&& \frac{dm}{dr}=4\pi r^2\rho,
\end{eqnarray}
where $G$ is the gravitational constant and $c$ is the speed of light.
Equations (\ref{TOVeq1}) and (\ref{TOVeq2}) are solved with the EoS
\begin{equation}\label{EoS}
  P(r)=P[\rho(r)],
\end{equation}
and with the boundary condition
\begin{equation}\label{BC}
  \rho(r=0)=\rho_C,
\end{equation}
where $\rho_C$ is the central mass density.
The integration is done on a uniform grid with $10^4$ points between $r=0$ and $r=2R_0$, where $R_0=10^6$ cm, for the following range of the central densities:
\begin{equation}\label{rhoCrange}
1.1\leq\frac{\rho_C}{\rho_0}\leq r_{max},
\end{equation}
where $r_{max}$ is a numerical factor.
The integration of Eqs. (\ref{TOVeq1}) and (\ref{TOVeq2}) stops once the matter density reaches the smallest value of $\rho$ available in the chosen EoS.
Following this procedure for each choice of $\rho_C$ one obtains the total mass $M$ and the radius $R$.
The EoS generally consists of three parts: for the crust, for the outer core and for the inner core.

\subsection{Core}
At present, it is generally not clear what are the effective degrees of freedom in the inner core (at densities above roughly $\rho_0$) and therefore, a reliable theoretical description of the inner core is not available.
However, one may build an envelope of equations of state constrained by the most general physical requirements \cite{HebelerEtAl2013} using the \emph{polytropic extension} of the EoS.
In this method, in the inner core, where $\frac{\rho}{\rho_0}>1.1$, the pressure $P[\rho(r)]$ is assumed to have the following form:
\begin{equation}\label{EoSinnerCore}
  P[\rho(r)]\propto\rho^\Gamma,
\end{equation}
where $\Gamma$ is a numerical factor with piecewise dependence on $\rho$ and with a condition that $P(r)$ is continuous everywhere inside the star.
Following Hebeler, Lattimer, Pethick and Schwenk \cite{HebelerEtAl2013} I use three polytropes which represent soft, intermediate and stiff variants of EoS in the inner core.
In addition, I use the BCPM EoS.

In the outer core with $0.5\leq\frac{\rho}{\rho_0}\leq1.1$, I use the parametrization of EoS proposed in \cite{HebelerEtAl2013} and in addition the BCPM EoS.
The internal energy density of {saturated} nuclear matter $\varepsilon(x,n)$ is given in equation (2) in \cite{HebelerEtAl2013}.
The total mass density of {saturated} nuclear matter including the electrons is
\begin{equation}\label{EoSOuterCore}
  \rho=m_pxn+m_n(1-x)n+\frac{n\varepsilon}{c^2}+\frac{(9\pi)^{2/3}}{4}\frac{\hbar}{c}\left(xn\right)^{\frac{4}{3}},
\end{equation}
where $m_p$ is the proton rest mass,
\begin{equation}\label{x}
  x=\frac{n_p}{n}
\end{equation}
is the ratio of the proton number density $n_p$ of saturated (uniform) nuclear matter divided by the baryon number density $n$ and $\varepsilon$ is the internal energy per baryon:
\begin{eqnarray}
 \nonumber && \varepsilon(n,x)=\varepsilon_0\left[\frac{3}{5}\left[x^{\frac{5}{3}} + \left(1-x\right)^{\frac{5}{3}}\right]\left(\frac{2n}{n_0}\right)^{\frac{2}{3}}\right.\\
 \label{e_x_n} &&\left.-\left[\alpha_1\left(x-x^2\right) + \alpha_2\right]\frac{n}{n_0}+\left[\eta_1\left(x-x^2\right) + \eta_2\right]\left(\frac{n}{n_0}\right)^\gamma \right].
\end{eqnarray}
Following \cite{HebelerEtAl2013} we have $\varepsilon_0=36.84$ MeV, $\alpha_1=2\alpha-4\alpha_L$, $\alpha_2=\alpha_L$, $\eta_1=2\eta-4\eta_L$, $\eta_2=\eta_L$.

It is useful to recall the general expansion of the energy per baryon in nearly symmetric nuclear matter in terms of the saturation energy $\omega_0$, incompressibility of symmetric nuclear matter $K_0$, the symmetry energy at saturation $S_0$ and its density derivative $L$ \cite{Lattimer1981,GonzalezBoquera2019}:
\begin{eqnarray}\label{wKSL}
 && \varepsilon(n,x)=-\omega_0 \\
\nonumber && +\frac{K_0}{18}\left(\frac{n}{n_0}-1\right)^2+\left[S_0+\frac{L}{3}\left(\frac{n}{n_0}-1\right)\right]\left(1-2x\right)^2.
\end{eqnarray}

From the empirical saturation properties of symmetric nuclear matter, namely the interaction energy per nucleon, $\varepsilon(n=n_0,x=1/2)=-\omega_0$, where
\begin{equation}\label{def_w0}
\omega_0=16\;{\rm MeV},
\end{equation}
and the nuclear saturation condition, $P_{\rm nuc}(n=n_0,x=1/2)=0$, it is easy to see that the parameters $\alpha$, $\eta$ and $\gamma$ in Eq. (\ref{e_x_n}) are related.
One finds
\begin{eqnarray}
\label{alpha}  && \alpha=\frac{4}{5} + \frac{2\gamma}{\gamma-1}\left(\frac{1}{5} + \frac{\omega_0}{\varepsilon_0}\right), \\
\label{eta}  && \eta=\frac{2}{\gamma-1}\left(\frac{1}{5} + \frac{\omega_0}{\varepsilon_0}\right).
\end{eqnarray}
The incompressibility $K=9n_0^2\partial^2\varepsilon/\partial n^2|_{n=n_0,\,x=1/2}$, the symmetry energy at saturation $S_0=\varepsilon(n_0,0)-\varepsilon(n_0,1/2)$ and the slope parameter of the symmetry energy $L=(3/8)n_0\partial_n\partial_x\partial_x\varepsilon|_{n=n_0,\,x=1/2}$ are given by
\begin{eqnarray}
\label{K} &&   K=9\varepsilon_0\left[-\frac{2}{15} + \gamma\left(\frac{1}{5} + \frac{\omega_0}{\varepsilon_0}\right)\right], \\
\label{S0} &&   S_0=\varepsilon_0\left( \frac{3}{5}2^{2/3} + \frac{\omega_0}{\varepsilon_0} - \alpha_L + \eta_L \right), \\
\label{L} &&   L=3\varepsilon_0\left( \frac{2}{5} - \alpha_L + \gamma\eta_L \right).
\end{eqnarray}
Equations (\ref{K})-(\ref{L}) relate the set of dimensionless parameters $(\gamma,\alpha_L,\eta_L)$ to the usual set $(K,S_0,L)$.
The parameters $\alpha_L$, $\eta_L$ with uncertainties are shown in figure 3 of \cite{HebelerEtAl2013}, and the parameter $\gamma$ is varied in the range $1.2\leq\gamma\leq1.45$.

In this paper I will use two sets of the parameters following \cite{HebelerEtAl2013}.
The first set is
\begin{equation}\label{set1}
(\gamma,\alpha_L,\eta_L)=(4/3,1.385,0.875),
\end{equation}
which leads to $(K,S_0,L)=(236\,{\rm MeV},32.3\,{\rm MeV},20.1\,{\rm MeV})$.
In some calculations which I will explicitly mention below, for comparison, I will use the second parameter set
\begin{equation}\label{set2}
(\gamma,\alpha_L,\eta_L)=(1.45,1.59,1.11),
\end{equation}
which leads to $(K,S_0,L)=(261\,{\rm MeV},33.4\,{\rm MeV},46.4\,{\rm MeV})$.

The partial nuclear pressure {(i.e. the total pressure in saturated nuclear matter excluding the electron pressure)} is
\begin{equation}\label{Pnuc}
  P_{\rm nuc}=n^2\left.\frac{\partial\varepsilon}{\partial n}\right|_x,
\end{equation}
and the total pressure (including the electron pressure) is
\begin{equation}\label{Pressure}
  P=P_{\rm nuc} + \frac{c\hbar}{4}(3\pi^2)^{1/3}(xn)^{4/3},
\end{equation}
where the second term accounts for the electron contribution.

The EoS in Eqs. (\ref{e_x_n}) and (\ref{Pressure}) is constrained by the ChEFT of nucleon interactions and is valid for uniform nuclear matter at densities $0.5\leq\frac{\rho}{\rho_0}\leq1.1$, while for the energy density $\frac{\rho}{\rho_0}>1.1$ the theoretical predictions are not reliable because there is not enough information on the relevant degrees of freedom \cite{HebelerEtAl2013}.
For $\frac{\rho}{\rho_0}\sim0.5$, theoretical predictions may be obtained by adding to the EoS contributions due to the nuclear surface energy \cite{BBP1971,WatanabeEtAl2000}.

Figure 1 shows the function $\varepsilon(x,n)$ for the first set of parameters (gold surface) and the corresponding nuclear energy per baryon $W(k,x)$ (green surface) introduced earlier by Baym, Bethe and Pethick (equation (3.19) in \cite{BBP1971}).
\begin{figure}
\includegraphics[width=3.5in]{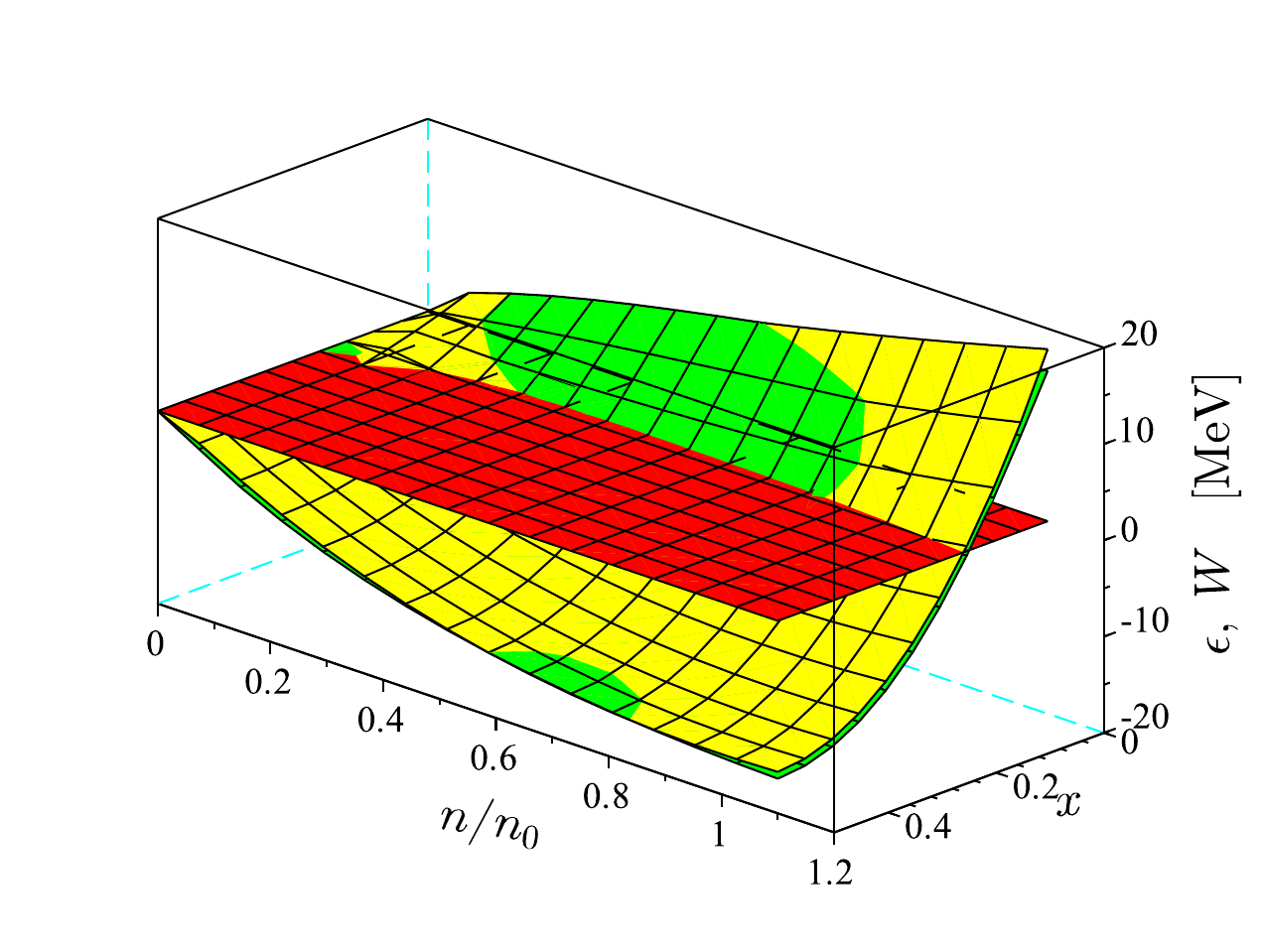}
\caption{Comparison of the nuclear energy per baryon $W(k,x)$ introduced by Baym, Bethe and Pethick (equation (3.19) in \cite{BBP1971}) and the nuclear energy per baryon $\varepsilon(x,n)$, Eq. (\ref{e_x_n}), with parameters defined in Eq. (\ref{set1}). The information provided by both baryon number density $n$ and proton fraction $x$ dependence allows to study the nuclear liquid crystal phases in the crust-core boundary. Red shows the zero surface, gold surface shows $\varepsilon(x,n)$ and green surface shows $W(k,x)$.}
\end{figure}
Although the interpolating formulas for $\varepsilon(n,x)$ and $W(k,x)$ look very different, it is seen from Fig. 1 that their predictions agree well.
The largest deviation of the two surfaces from one another is 2.23 MeV at $(n,x)=(1.1n_0,0)$.
In case when the second set of parameters defined in Eq. (\ref{set2}) is used the resulting surfaces still look similar but the maximum deviation of surfaces is seen again at $(n,x)=(1.1n_0,0)$ of the order of 4.3 MeV.

Figures 2 and 3 show different representative EoS in the inner core and in the outer core, correspondingly.
In addition to the soft, intermediate and stiff representative EoS, the plots show also the predictions of the BCPM EoS presented by Sharma, Centelles, Vinas, Baldo and Burgio in \cite{SharmaEtAl2015}.

The data in the outer core was calculated for the representative EoS, from Eq. (\ref{EoSOuterCore}), with the beta-equilibrium condition,
\begin{equation}\label{betaEq}
  m_pc^2+\left.\frac{\partial (n\varepsilon)}{\partial n_p}\right|_{n_n} + \mu_e=m_nc^2+\left.\frac{\partial (n\varepsilon)}{\partial n_n}\right|_{n_p},
\end{equation}
where $n_p=xn$ and $n_n=(1-x)n$ is the number density of protons and neutrons inside the saturated nuclear matter, respectively.
{The electron chemical potential is given by
\begin{equation}\label{def_mue}
  \mu_e=\hbar c(3\pi^2 xn)^{1/3},
\end{equation}
and the baryon chemical potentials are given by
\begin{eqnarray}
\label{mun} && \mu_n(n,x)=\left( \left.\frac{\partial }{\partial n}\right|_{x} - \frac{x}{n}\left.\frac{\partial }{\partial x}\right|_{n} \right)(n\varepsilon(n,x)),  \\
\label{mup} && \mu_p(n,x)=\left( \left.\frac{\partial }{\partial n}\right|_{x} + \frac{1-x}{n}\left.\frac{\partial }{\partial x}\right|_{n} \right)(n\varepsilon(n,x)).
\end{eqnarray}
}

\begin{figure}
\includegraphics[width=3.5in]{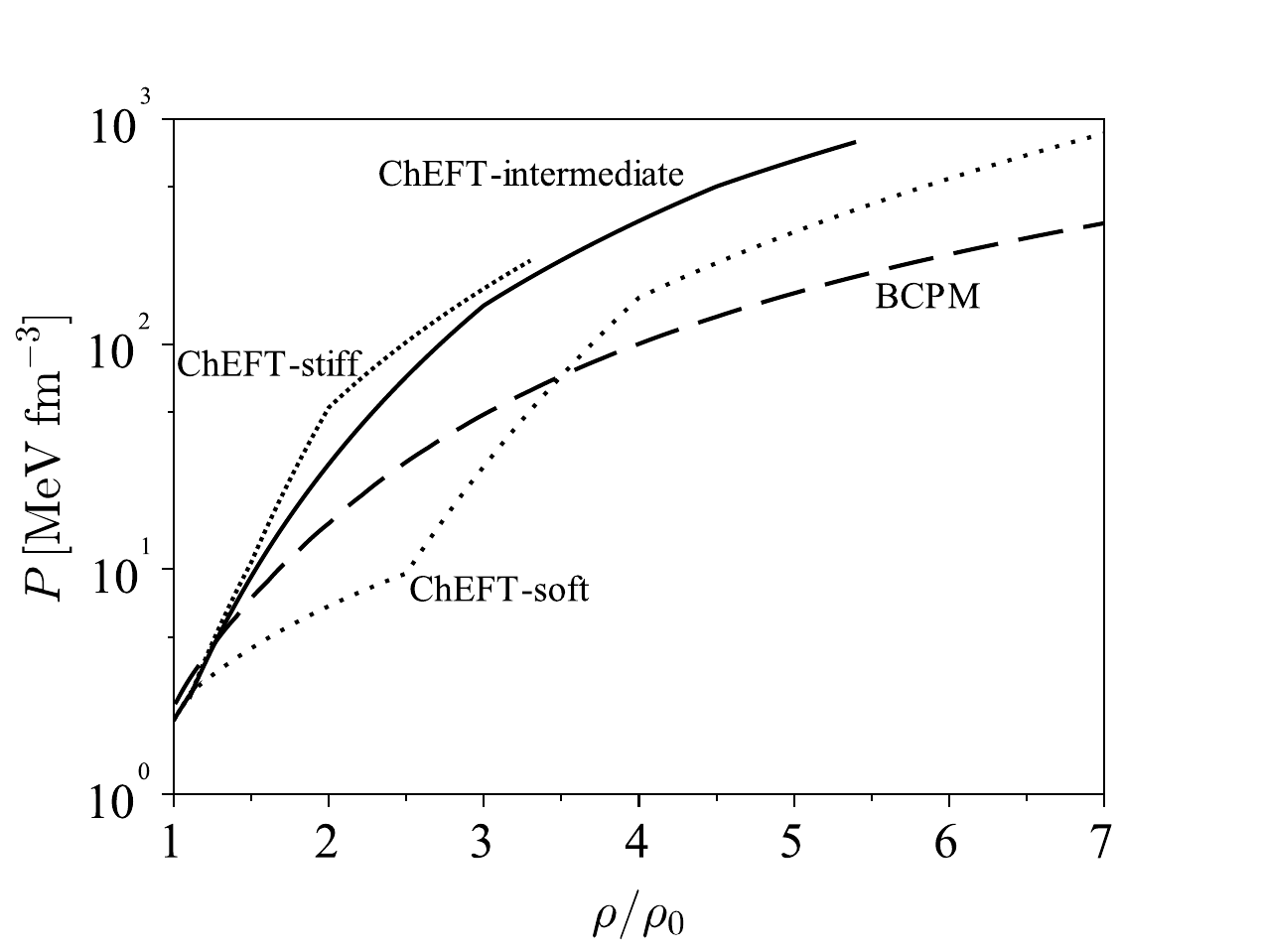}
\caption{The EoS in the inner core.  Pressure calculated from three representative ChEFT EoS with the parameters $(\gamma,\alpha_L,\eta_L)$ from Eq. (\ref{set1}) and extrapolated by the general polytropic extension \cite{HebelerEtAl2013}. Dashed line shows pressure calculated from BCPM EoS \cite{SharmaEtAl2015}. The reference energy density of saturated nuclear matter is $\rho_0=2.68\times10^{14}\;{\rm g\;cm^{-3}}$. The polytropic indices are: $\Gamma=1.5$ for $1.1\leq\frac{\rho}{\rho_0}\leq2.5$, $\Gamma=6$ for $2.5\leq\frac{\rho}{\rho_0}\leq4$, $\Gamma=3$ for $4\leq\frac{\rho}{\rho_0}\leq7$ -- for the soft EoS;
$\Gamma=4$ for $1.1\leq\frac{\rho}{\rho_0}\leq3$, $\Gamma=3$ for $3\leq\frac{\rho}{\rho_0}\leq4.5$, $\Gamma=2.5$ for $4.5\leq\frac{\rho}{\rho_0}\leq5.4$ -- for the intermediate EoS;
$\Gamma=4.5$ for $1.1\leq\frac{\rho}{\rho_0}\leq1.5$, $\Gamma=5.5$ for $1.5\leq\frac{\rho}{\rho_0}\leq2$, $\Gamma=3$ for $2\leq\frac{\rho}{\rho_0}\leq3.3$ -- for the stiff EoS.}
\end{figure}
\begin{figure}
\includegraphics[width=3.5in]{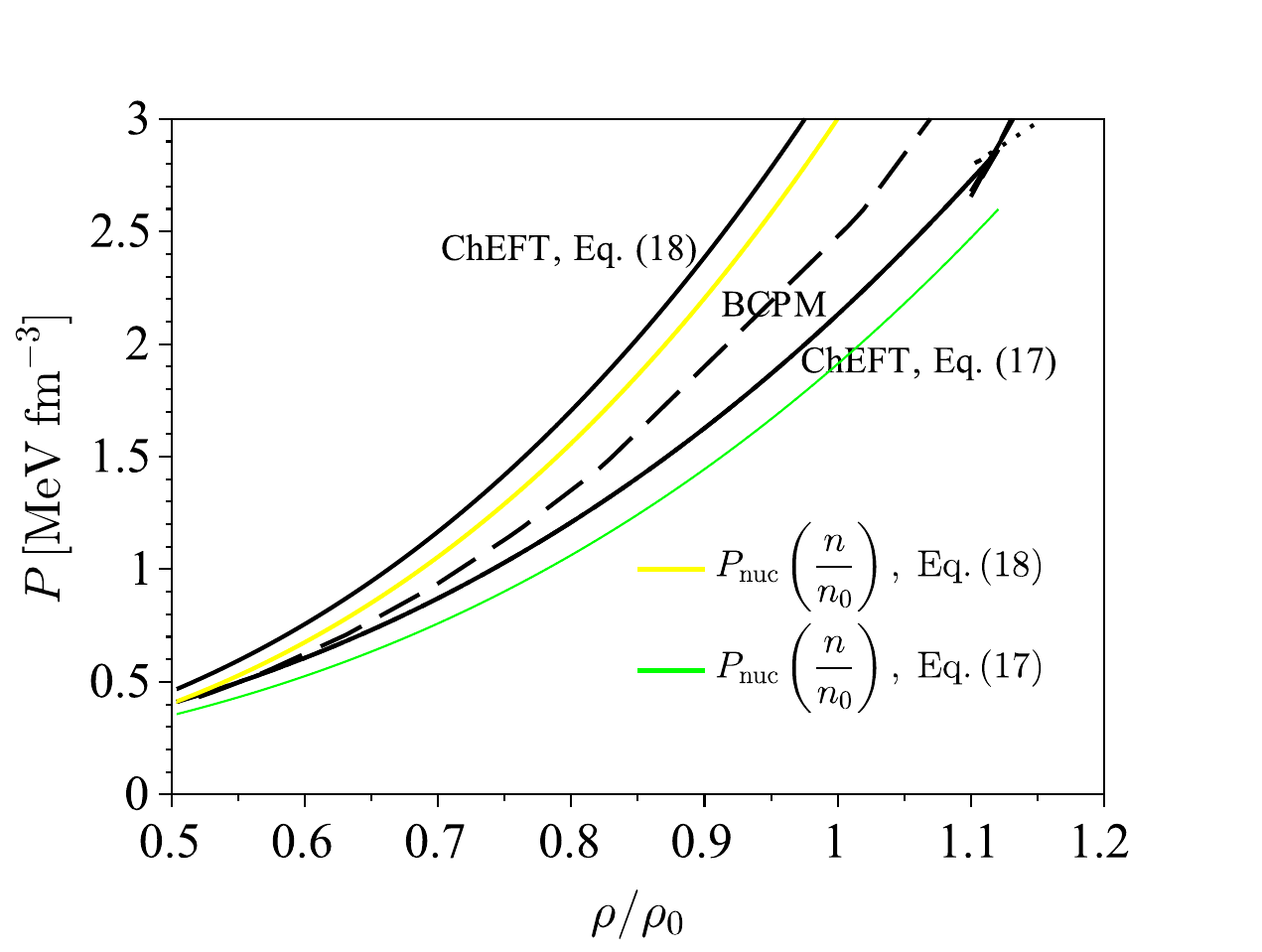}
\caption{The EoS in the outer core. Solid black lines show pressure $P$ calculated from ChEFT EoS with the parameters $(\gamma,\alpha_L,\eta_L)$ from Eq. (\ref{set1}) and Eq. (\ref{set2}). Dashed line shows pressure $P$ calculated from BCPM EoS \cite{SharmaEtAl2015}. Gold and green lines show the partial nuclear pressure $P_{\rm nuc}$ (equal to $P$ minus the electron contribution) in the outer core. Figure shows $P$ as a function of $\rho/\rho_0$ and $P_{\rm nuc}$ as a function of $n/n_0$.}
\end{figure}

The data for BCPM is taken from table 9 of \cite{SharmaEtAl2015} and one can note that these data are identical to the data given in table 5 of the work by Douchin and Haensel \cite{DouchinHaensel2001}.
It is seen that the chosen EoS provide rather similar predictions in the outer core, but in the inner core the differences are notable.
It turns out that at higher densities, $\frac{\rho}{\rho_0}\geq3.5$, the BCPM data represents an even softer EoS than the soft representative EoS generated by the polytropic extension.

A notable practical advantage of the ChEFT-EoS given in Eq. (\ref{e_x_n}) is that it contains information on the nuclear energy for arbitrary values of proton fraction $x$ for given baryon number density $n$ and thus gives semi-analytical access to physics on the crust-core boundary as will be shown below.

\subsection{Crust}
The crust is characterized by the baryon densities $\frac{\rho}{\rho_0}\lesssim1.7\times10^{-3}$ for the outer crust and $1.7\times10^{-3}\lesssim\frac{\rho}{\rho_0}\lesssim0.5$ for the inner crust.
The corresponding data is plotted in Fig. 3 by dashed line, where the numerical values are taken from tables 4 and 7 reported by Sharma, Centelles, Vinas, Baldo and Burgio in \cite{SharmaEtAl2015}.
For comparison, the results reported by Lattimer and Swesty in \cite{LattimerSwesty1991} are shown by solid line.
Figure 4 shows the both sets of data on the same plot and an excellent agreement in the inner crust is seen.
\begin{figure}
\includegraphics[width=3.5in]{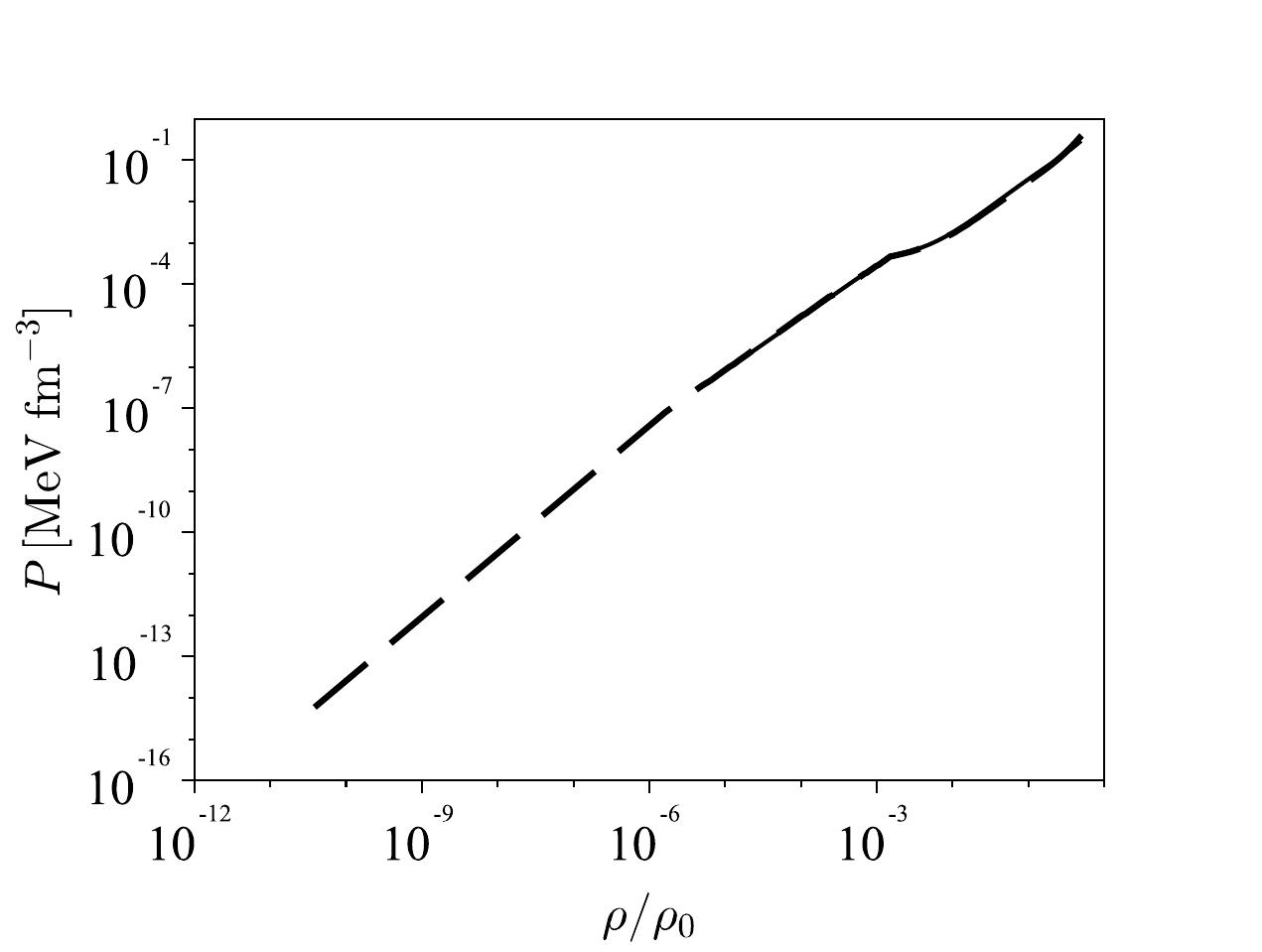}
\caption{The EoS in the crust. The solid line shows data found by Lattimer and Swesty in \cite{LattimerSwesty1991} for the inner crust. The dashed line shows the data from BCPM EoS \cite{SharmaEtAl2015} for both the inner crust and the outer crust.}
\end{figure}

\subsection{Crust-core boundary}
At the crust-core boundary, one expects that the nuclei are nonspherical -- the so called pasta phases \cite{RavenhallEtAl1983,HashimotoEtAl1984}.
The density range and the composition of the pasta phases appear to be very sensitive on a model used to describe the nuclear energy.
Physics of the nonspherical nuclei in the pasta phases can be understood as a competition of the surface and the {Coulomb} energies.
Detailed description of the crust can be obtained by minimizing the total energy density with respect to a set of variables
\begin{equation}\label{setVar}
n,\;x,\;n_{no},\;r_N,\;u,
\end{equation}
where $n_{no}$ is the number density of neutrons outside (in between) nuclei, $r_N$ is the radius of a nucleus and $u$ is the volume fraction of nucleus in the Wigner-Seitz cell, as explained in \cite{BBP1971} and \cite{WatanabeEtAl2000}.
Recently, an explosive growth of calculations based on various EoS which reveal presence of the pasta phases has emerged, however the uncertainties remain large.
At present, it is important to find additional theoretical constraints and computational tests that would help to sort out predictions of various models for the pasta phases.

In order to reveal basic physics of the pasta phases one needs to know (i) the nuclear energy per baryon $\varepsilon(n,x)$ for pairs of variables $(n,x)$ out of beta equilibrium, (ii) the nuclear surface energy, (iii) the Coulomb energy including the self energy of nuclei and the interaction energy between nuclei and between nuclei and the electrons.

A quantity of practical importance is the proton chemical potential in the pure neutron phase, $\mu_{po}$, and the difference between $\mu_{po}$ and the proton chemical potential in the saturated nuclear matter, $\mu_{pi}$, at the same pressure as the pure neutron phase.
In order to make a contact with the work by Watanabe, Iida and Sato \cite{WatanabeEtAl2000}, I notice that the parametrization of $\mu_{po}$ has been suggested in equation (4) of \cite{WatanabeEtAl2000},
\begin{equation}\label{mu_p_0_2000}
  \mu_{po}=-C_1n_{no}^{2/3},
\end{equation}
where $C_1$ is a positive definite numerical factor.
Watanabe, Iida and Sato consider for $C_1$ the values 300, 400 and 600 ${\rm MeV\;fm^2}$.
From the ChEFT EoS adopted in the present paper, Eq. (\ref{e_x_n}), using Eq. (\ref{mup}) with $x=0$ I obtain
\begin{equation}\label{mu_p_0}
  \mu_{po}=\varepsilon_0\left[\left(\frac{n_{no}}{n_0}\right)^{\gamma}\left[\eta_1+(1+\gamma)\eta_2\right]-\frac{n_{no}}{n_0}(\alpha_1+2\alpha_2)\right].
\end{equation}

\begin{figure}
\includegraphics[width=3.5in]{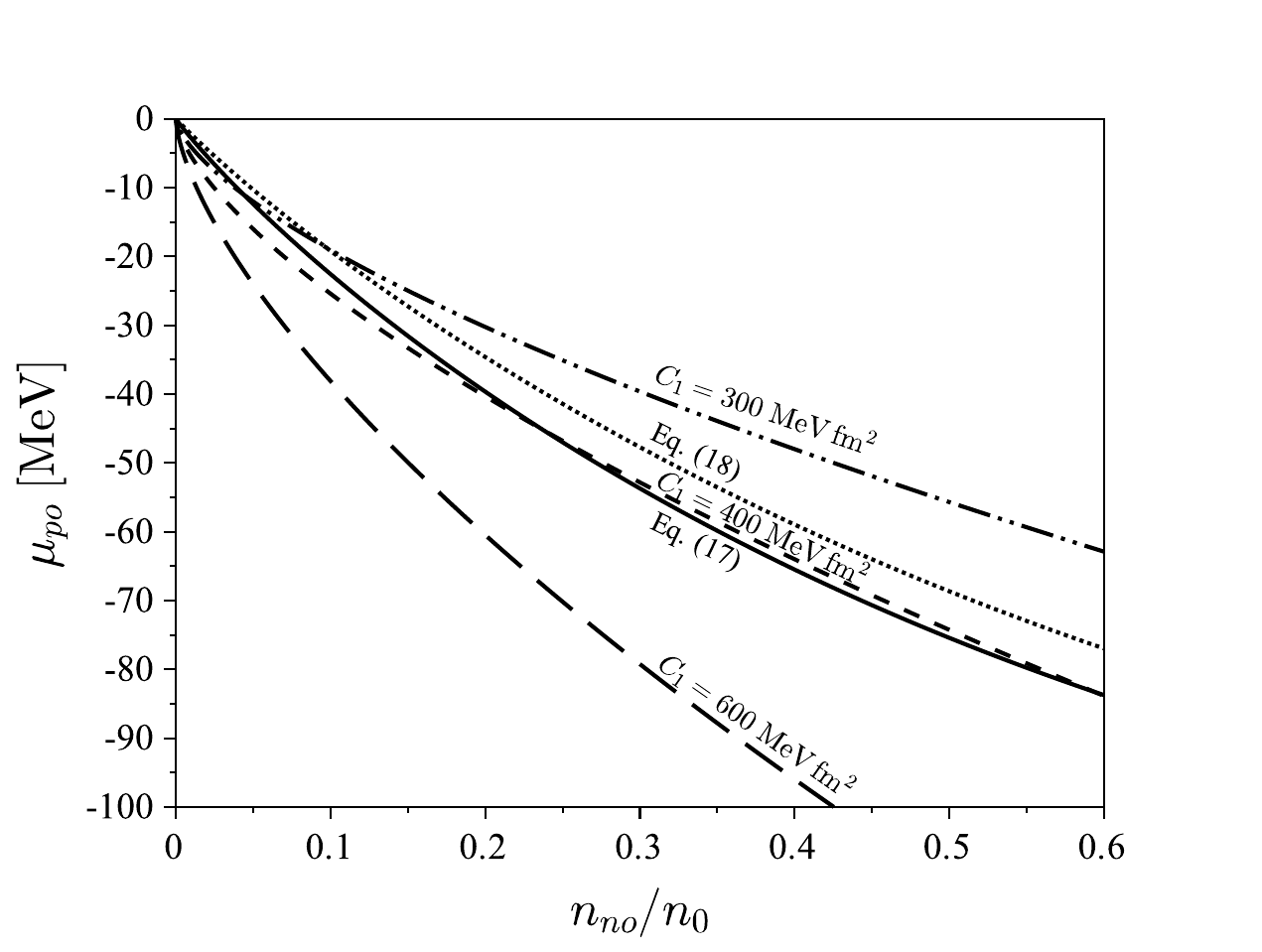}
\caption{The proton chemical potential in the pure neutron matter $\mu_{po}$ calculated from the ChEFT  \cite{HebelerEtAl2013} obtained from Eq. (\ref{mup}) with $x=0$, with the parameters $(\gamma,\alpha_L,\eta_L)$ from Eq. (\ref{set1}) (solid line) and in Eq. (\ref{set2}) (dotted line). Dash-dotted, short-dashed and long-dashed lines show three different choices of the parameter $C_1$ in $\mu_{po}$ used by Watanabe, Iida and Sato in \cite{WatanabeEtAl2000}.}
\end{figure}
Figure 5 shows the proton chemical potential in the pure neutron phase calculated with parameterizations suggested in \cite{WatanabeEtAl2000} and in \cite{HebelerEtAl2013} with two parameter sets defined in Eqs. (\ref{set1}) and (\ref{set2}).

The liquid crystal nature of the inner crust is manifested already in the phase with spherical nuclei because of the dripped neutron outside of nuclei.
An ordinary nucleus has zero pressure and is in equilibrium with vacuum.
In contrast, the pure neutron matter has a positive pressure and can be in equilibrium only if a positive pressure is supported by the nucleus, which thus becomes neutron-rich.
Neutron-rich nuclei in the inner crust are in equilibrium with the pure neutron matter and thus the pressures in the both phases are equal.
Another equilibrium condition is that the neutron chemical potential in the both phases are equal.

Thus, the equilibrium proton fraction inside a neutron-rich nucleus can be found by calculating, at a given partial nuclear pressure $P_{\rm nuc}$, the neutron chemical potential $\mu_n$ as function of the proton fraction $x$ and finding the nonzero value of $x$ at which the resulting $\mu_n$ is equal to $\mu_n$ at $x=0$.
{With the nuclear energy $W(k,x)$ and neglecting the surface and the Coulomb energy, such calculation was done by Baym, Bethe and Pethick in \cite{BBP1971} and the results were displayed in figure 2 of their paper.}
{In this paper, with the nuclear energy $\varepsilon(n,x)$, Eq. (\ref{e_x_n}) with the parameters $(\gamma,\alpha_L,\eta_L)$ from Eq. (\ref{set1}) and neglecting the surface and the Coulomb energy, the neutron chemical potential $\mu_n$ as function of the proton fraction $x$ at fixed partial nuclear pressure $P_{\rm nuc}$ is calculated in this paper  and the result for $\gamma=4/3$ is shown in Fig. 6. }
The plot is analogous to figure 2 in \cite{BBP1971} except that $P_{\rm nuc}$ is calculated from the modern EoS given in Eq. (\ref{e_x_n}).
\begin{figure}
\includegraphics[width=3.5in]{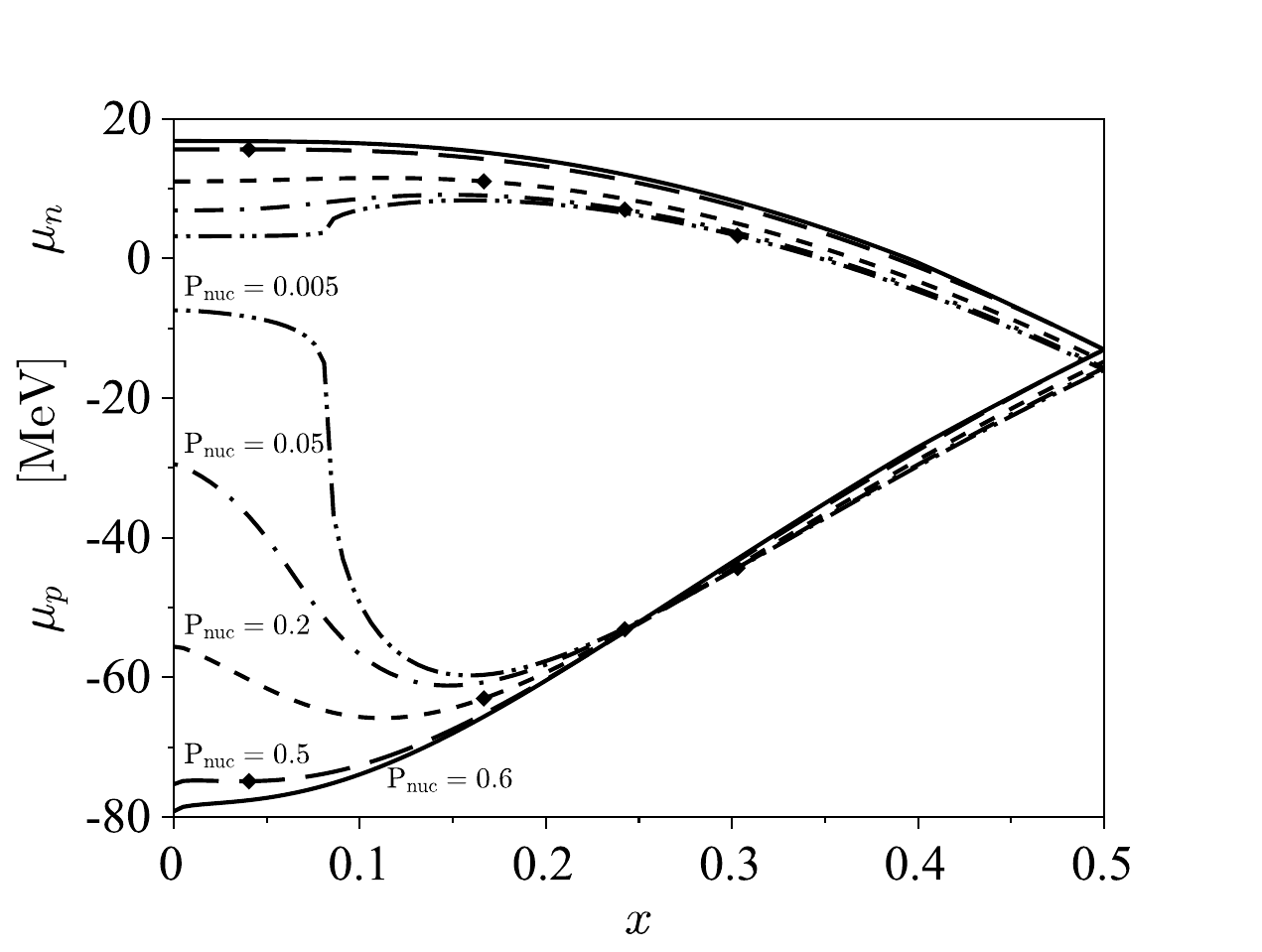}
\caption{The neutron $\mu_n$ and proton $\mu_p$ chemical potentials, Eqs. (\ref{mun}) and (\ref{mup}) for bulk nuclear matter, plotted for given partial nuclear pressures $P_{\rm nuc}$ as a function of $x$, constrained by beta-equilibrium.
ChEFT EoS is used with Eq. (\ref{set1}). Each type of the lines correspond to different values of $P_{\rm nuc}$.
The curves $\mu_n$ start with $x=0$ at the positive values and meet at $x=1/2$ with the curves $\mu_p$.
The rhombuses on $\mu_n$ mark the coexistence of pure neutron matter and the saturated nuclear matter defined by equality of $\mu_n$ in pure neutron matter and in the saturated matter; the corresponding values of $\mu_p$ in the saturated matter are marked by rhombuses on the curves $\mu_p$.
Rhombuses on $\mu_n$ disappear for $P_{\rm nuc}>0.5\;{\rm MeV\,fm^{-3}}$ which implies that the coexistence becomes impossible.
Rhombuses on $\mu_p$ help to see magnitude of the contact difference of the proton chemical potentials $\Delta\mu_p$, Eq. (\ref{Deltamup}), in the nucleus and the pure neutron matter.}
\end{figure}
Each of the curves starting at positive (or negative) values of the vertical axis corresponds to the neutron $\mu_n$ (or proton $\mu_p$) chemical potential at fixed pressure $P_{\rm nuc}$.

In Fig. 6, the black filled rhombuses on the $\mu_n$ curves mark the points where the value of $\mu_n$ at nonzero $x$ is equal to the value of $\mu_n$ at $x=0$ on the same curve for given fixed $P_{\rm nuc}$; for this value of $P_{\rm nuc}$, there is a corresponding curve of $\mu_p$ and a rhombus that marks the value of $\mu_p$ at the coexistence of nucleus and the pure neutron matter.
It should be noted that the coexistence also includes the condition that the value of the proton chemical potential $\mu_p$ marked by the rhombus be more negative than $\mu_p$ at $x=0$ on the same curve for given fixed $P_{\rm nuc}$; this implies that it is energetically favorable for the protons to stay inside the nucleus.
Figure 6 shows the data with the assumption that the surface and the Coulomb energy contributions may be neglected.
This assumption is expected to be good enough near the crust-core transition.

\subsection{{Liquid drop model}}
{A better quantitative picture can be achieved from the elementary considerations in the framework of the compressible liquid drop model.
In this model, the system is assumed to be in the ordered state consisting of elementary Wigner-Seitz cells.
In the region of spherical nuclei ($d=3$), the Wigner-Seitz cell may be assumed spherical, in the region of rod-like nuclei ($d=2$) it is a circle and in the region of slab-like nuclei ($d=1$) it is a line segment.
I use $E$ to express the energy and $w$ to express the energy density.
The volume of the Wigner-Seitz cell is $V_c=(4\pi/3)r_c^3$ for $d=3$, $V_c=\pi r_c^2\times l\,{\rm cm}$ for $d=2$ and $V_c=r_c\times l^2\,{\rm cm}^2$ for $d=1$, where $l\rightarrow+\infty$ is the size of the system along the uniform direction.
The volume fraction of the dense nuclear matter is equal to
\begin{equation}\label{def_u}
  u=\left(\frac{r_N}{r_c}\right)^d.
\end{equation}
The baryon number density averaged across the Wigner-Seitz cell is
\begin{equation}\label{def_nb}
  n_b=un+(1-u)n_{no}.
\end{equation}
The total energy $E_{\rm tot}$ in a single Wigner-Seitz cell includes: the energy density of nucleus $w_{\rm nuc}$ excluding the self Coulomb energy and including the rest mass density, the surface energy density $w_{\rm surf}$, the Coulomb energy density including the lattice energy density $w_{\rm C+L}$, energy density of the dripped neutrons $w_{no}$ and the electron energy density $w_{\rm e}$ and is given by
\begin{equation}\label{def_Etot}
  E_{\rm tot}(n,x,n_{no},r_N,u)=V_c\left[w_{\rm nuc} + w_{\rm s} + w_{\rm C+L} + w_{no} + w_{\rm e}\right],
\end{equation}
with
\begin{eqnarray}
\nonumber &&  w_{\rm nuc}=un\left[\left(1-x\right)m_n + xm_p\right]c^2 + un\varepsilon(n,x), \\
   \\
\nonumber &&  w_{\rm s}=w_{\rm s}(n,x,n_{no},r_N,u), \\
\nonumber &&  w_{\rm C+L}=w_{\rm C+L}(xn,r_N,u), \\
\nonumber &&  w_{no}=\left(1-u\right)n_{no}\left[m_nc^2 + \varepsilon(n_{no},0)\right], \\
\nonumber &&  w_{\rm e}=\frac{3}{4}\hbar c(3\pi^2)^{1/3} (unx)^{4/3},
\end{eqnarray}
where $n=n_{pi}+n_{ni}$ is the baryon number density inside the nucleus, $\varepsilon(n,x)$ is the energy per baryon in the dense matter phase (inside the nucleus), $x$ is the proton fraction in the dense matter phase
\begin{equation}\label{def_x}
  x=\frac{n_{pi}}{n},
\end{equation}
$n_{no}$ is the number density of neutrons outside the nucleus and in writing of $w_{\rm e}$ the charge neutrality condition in a single Wigner-Seitz cell has been used.

To describe the surface energy I follow Watanabe, Iida and Sato \cite{WatanabeEtAl2000} and use the following expression:
\begin{equation}\label{def_wsurf}
  w_{\rm s}=\frac{ud}{r_N}C_2\tanh\left[\frac{C_3}{\mu_n^{(0)}}\right]\frac{\sigma\left(n-n_{no}\right)^{2/3}}{(36\pi)^{1/3}\omega_0}\left[\varepsilon(n_{no},0)-\varepsilon(n,x)\right].
\end{equation}
The standard choice \cite{WatanabeEtAl2000} for the parameters is: $C_2=1$ MeV, $C_3=3.5$ MeV, $\sigma=21.0$ MeV and $\omega_0$ is given by Eq. (\ref{def_w0}).
Here the neutron chemical potential excluding the surface corrections and the rest mass is
\begin{equation}\label{mun0}
  \mu_n^{(0)}=\frac{1}{1-u}\frac{dw_{no}}{dn_{no}}-m_nc^2.
\end{equation}

Actually, the surface energy density $w_{\rm s}$ in Eq. (\ref{def_wsurf}) has been derived from the EoS parameterized by $W(k,x)$.
The requirement of self-consistency is that $w_{\rm s}$ is derived from the same EoS as is used for the description of the dense phase.
This requirement is fulfilled in \cite{WatanabeEtAl2000}, but in this paper $w_{\rm s}$ must have been derived from $\varepsilon(n,x)$.
However, for order of magnitude estimates needed in this paper it is sufficient to use the form in Eq. (\ref{def_wsurf}) with $\varepsilon(n,x)$.

{Finally, the Coulomb energy density including the self Coulomb and the lattice energies is given by
\begin{equation}\label{def_wCoul}
  w_{\rm C+L}=2\pi(enxr_N)^2uf_d(u),
\end{equation}
where \cite{RavenhallEtAl1983}
\begin{equation}\label{def_fdu}
  f_d(u)=\frac{1}{d+2}\left[\frac{2}{d-2}\left(1-\frac{du^{1-2/d}}{2}\right) + u\right],
\end{equation}
with \cite{PethickRavenhall1995}
\begin{eqnarray}
\nonumber && f_3(u)=\frac{1}{5}(2-3u^{1/3}+u), \\
\nonumber && f_2(u)=\frac{1}{4}\left(\ln\frac{1}{u}-1+u\right), \\
\nonumber && f_1(u)=\frac{1}{3}\left(\frac{1}{u}-2+u\right).
\end{eqnarray}
}

{The equilibrium nuclear shapes can be found by minimization of $E_{\rm tot}$ with respect to:
\begin{itemize}
  \item $r_N$ at fixed $n$, $x$, $n_{no}$ and $u$,
  \item $n_{ni}$ at fixed $N_{n}$, $n_{pi}$, $r_N$ and $u$, where $N_n$ is the total number of neutrons in the Wigner-Seitz cell,
  \item $x$ at fixed $n$, $n_{no}$, $r_N$ and $u$ and
  \item $u$ at fixed $N_{n}$, $N_{no}$, $x$ and $V_c$, where $N_{no}$ is the number of neutrons outside the nucleus.
\end{itemize}}

{Minimization of $E_{\rm tot}$ with respect to $r_N$ (at fixed $n$, $x$, $n_{no}$ and $u$) leads to the well-known relation
\begin{equation}\label{EquilCond_r}
  w_{\rm s}=2w_{\rm C+L}.
\end{equation}
}

{Minimization of $E_{\rm tot}$ with respect to $n_{ni}$ (at fixed $N_{n}$, $n_{pi}$, $r_N$ and $u$) expresses the continuity of the neutron chemical potential across the dense matter phase and the pure neutron matter phase.
Fixing $N_{n}$, $n_{pi}$, $r_N$ and $u$ induces the following relations: $n\delta x + x\delta n=0$ and $u\delta n_{ni} + (1-u)\delta n_{no}=0$.
Therefore,
\begin{eqnarray}\label{partial_nni}
 &&  \left.\frac{\partial}{\partial n_{ni}}\right|_{N_{n},\,n_{pi},\,r_N,\,u}= \\
  \nonumber && \left.\frac{\partial}{\partial n}\right|_{x,\,n_{no},\,r,\,u} - \frac{x}{n}\left.\frac{\partial}{\partial x}\right|_{n,\,n_{no},\,r_N,\,u} - \frac{u}{1-u}\left.\frac{\partial}{\partial n_{no}}\right|_{n,\,x,\,r_N,\,u}.
\end{eqnarray}
Application of these partial derivative yields
\begin{equation}\label{continuityMu}
  \mu_{ni}=\mu_{no},
\end{equation}
where
\begin{eqnarray}
\nonumber   \mu_{ni}=m_nc^2 + \left(\left.\frac{\partial}{\partial n}\right|_{x} - \frac{x}{n}\left.\frac{\partial}{\partial x}\right|_{n}\right)\left[n\varepsilon(n,x)\right] \\
\label{Muni}   + \frac{1}{u} \left.\frac{\partial  w_{\rm s}}{\partial n_{ni}}\right|_{N_{n},\,n_{pi},\,r_N,\,u},
\end{eqnarray}
and
\begin{equation}\label{Muno}
  \mu_{no}=\left.\frac{\partial  w_{no}}{\partial n_{no}}\right|_{u} + \frac{1}{1-u}\left.\frac{\partial w_{\rm s}}{\partial n_{no}}\right|_{n,\,x,\,r_N,\,u}.
\end{equation}
Notice that Eqs. (\ref{Muni}) and (\ref{Muno}) do not contain the contribution from $w_{\rm C+L}$ because ${\partial w_{\rm C+L}}/{\partial n_{ni}}|_{N_{n},\,n_{pi},\,r_N,\,u}=0$.
}

{Minimization of $E_{\rm tot}$ with respect to $x$ at fixed $n$, $n_{no}$, $r_N$ and $u$ leads to the beta-equilibrium condition:
\begin{eqnarray}
 &&  \label{betaEquilCond}
  \mu_e+(m_p-m_n)c^2 = \\
\nonumber  && -\left.\frac{\partial \varepsilon}{\partial x}\right|_{n} - \frac{1}{un}\left.\frac{\partial (w_{\rm s}+w_{\rm C+L})}{\partial x}\right|_{n,n_{no},r_N,u}.
\end{eqnarray}
}

{Finally, minimization of $E_{\rm tot}$ with respect to $u$ at fixed $N_{n}$, $N_{no}$, $x$ and $V_c$ expresses the continuity of the partial nuclear pressure across the dense matter phase and the pure neutron matter phase.
Fixing $N_{n}$, $N_{no}$, $x$ and $V_c$ induces the following relations: $n\delta u + u\delta n=0$ and $(1-u)\delta n_{no} - n_{no}\delta u=0$.
Therefore,
\begin{eqnarray}\label{partial_nni}
 &&  \left.\frac{\partial}{\partial u}\right|_{N_{n},\,N_{no},\,x,\,V_c}= \\
  \nonumber && \left.\frac{\partial}{\partial u}\right|_{n,\,x,\,n_{no}} - \frac{n}{u}\left.\frac{\partial}{\partial n}\right|_{n_{no},\,r_N,\,u} + \frac{n_{no}}{1-u}\left.\frac{\partial}{\partial n_{no}}\right|_{n,\,x,\,r_N,\,u}.
\end{eqnarray}
Application of these partial derivative yields
\begin{equation}\label{continuityP}
  P_{i}=P_{o},
\end{equation}
\begin{equation}\label{Pi}
  P_{i} = n^2\left.\frac{\partial \varepsilon}{\partial n}\right|_{x} - \left(\left.\frac{\partial}{\partial u}\right|_{n,\,x,\,n_{no}} - \frac{n}{u}\left.\frac{\partial}{\partial n}\right|_{n_{no},\,r_N,\,u}\right)(w_{\rm s} + w_{\rm C+L}),
\end{equation}
and
\begin{equation}\label{Po}
  P_{o} = \frac{n_{no}}{1-u}\left.\frac{\partial w_{no}}{\partial n_{no}}\right|_{u} - \frac{w_{no}}{1-u} + \frac{n_{no}}{1-u} \left.\frac{\partial  w_{\rm s}}{\partial n_{no}}\right|_{n,\,x,\,r_N,\,u}.
\end{equation}
}

{Equipped with these formulas it is straightforward to evaluate the magnitude of the surface and Coulomb contributions of the partially saturated nuclear matter (in the Wigner-Seitz cell) relative to the nuclear contribution of entirely saturated matter (in uniform nuclear matter).
Comparing Eqs. (\ref{mun}), (\ref{Muni}) and (\ref{Muno}) I obtain that the surface plus Coulomb corrections to chemical potentials inside the nucleus $[\mu_{ni}]$ and outside the nucleus $[\mu_{no}]$ are given by
\begin{eqnarray}
 \label{correctionsMu1} && [\mu_{ni}] = \frac{1}{u} \left.\frac{\partial  w_{\rm s}}{\partial n_{ni}}\right|_{N_{n},\,n_{pi},\,r_N,\,u},\\
 \label{correctionsMu2} && [\mu_{no}] = \frac{1}{1-u} \left.\frac{\partial  w_{\rm s}}{\partial n_{no}}\right|_{n,\,x,\,r_N,\,u}.
\end{eqnarray}
Comparing Eqs. (\ref{Pnuc}), (\ref{Pi}) and (\ref{Po}) I obtain that the surface plus Coulomb corrections to pressure inside the nucleus $[P_{i}]$ and outside the nucleus $[P_{o}]$ are given by
\begin{eqnarray}\label{correctionsP}
 && [P_{i}] = -\left(\left.\frac{\partial}{\partial u}\right|_{n,\,x,\,n_{no}} - \frac{n}{u}\left.\frac{\partial}{\partial n}\right|_{n_{no},\,r_N,\,u}\right)(w_{\rm s} + w_{\rm C+L}),\\
 && [P_{o}] = \frac{n_{no}}{1-u} \left.\frac{\partial  w_{\rm s}}{\partial n_{no}}\right|_{n,\,x,\,r_N,\,u}.
\end{eqnarray}
Using these formulas with $u=0.5$, $n_{no}=n_0/2$ and $x=0.05$, which roughly corresponds to typical values in the slab phase it is easy to find that the surface corrections are indeed small compared with the values obtained without those corrections.
For instance, from Eq. (\ref{correctionsMu2}) I find that $[\mu_{no}]\simeq-0.1$ MeV, which is small compared with $\mu_{no}\simeq15$ MeV as has been anticipated.
This confirms the expectation that the calculations of $\mu_n$ and $\mu_p$ displayed in Fig. 6 provide a reliable and rather precise picture of the coexistence of the dense and the pure neutron phases.
}

\section{Numerical results}
\subsection{Crust-core boundary}
Starting from Fig. 6 I investigate the equilibrium between the pure neutron phase and the saturated nuclear matter phase.
Comparing Fig. 6 based on $\varepsilon(n,x)$ given in Eq. (\ref{e_x_n}), with figure 2 from the work by Baym, Bethe and Pethick in \cite{BBP1971} based on the nuclear energy $W(k,x)$, it is clearly seen that in \cite{BBP1971} there is a noticeable maximum in the curve $\mu_n$ for $W(k,x)$ even at the pressure $P_{\rm nuc}=1.1\;{\rm MeV\,fm^{-3}}$, while the coexistence curve (the multitude of rhombuses on the $\mu_n$ curves) in Fig. 6 terminates for $P_{\rm nuc}>0.5\;{\rm MeV\,fm^{-3}}$.
The termination of the coexistence curve can be characterized by disappearance of the local maximum on the curve $\mu_n$ at the threshold value $P_{*}$ of $P_{\rm nuc}$.
For $P_{\rm nuc}>P_{*}$, the neutron chemical potential $\mu_n$ is always larger in the pure neutron phase than in the saturated nuclear matter phase and $\mu_p$ outside nucleus is always lower than $\mu_p$ inside the nucleus.
It is seen from Fig. 6 that the termination of the coexistence curve is accompanied by change of the sign of the contact difference of the proton chemical potentials in the nucleus and the pure neutron matter.
I notice that the present result $P_{*}\sim0.5\;{\rm MeV\,fm^{-3}}$ is similar to the result due to Centelles, Del Estal and Vinas shown in their figure 3 in \cite{Vinas1998}, where the pressure at coexistence is seen to be always below $0.5\;{\rm MeV\,fm^{-3}}$.

I turn to examination of the dependence of $P_{*}$ on the parameters $(\gamma,\alpha_L,\eta_L)$ in Eq. (\ref{e_x_n}).
I choose 100 equally spaced points that sample the values of $(\gamma,\alpha_L,\eta_L)$ between the two sets defined in Eqs. (\ref{set1}) and (\ref{set2}) according to
\begin{eqnarray}
  \nonumber &&(\gamma,\alpha_L,\eta_L)(j)=(4/3,1.385,0.875) \\
 \label{points100} && + \frac{j-1}{99}\left[(1.45,1.59,1.11) - (4/3,1.385,0.875)\right]
\end{eqnarray}
for $j$ between 1 and 100, where $(\gamma,\alpha_L,\eta_L)(j)$ implies the $j$-th choice for the value of $(\gamma,\alpha_L,\eta_L)$, which is treated as a row vector.
The computation is done in the following intervals of values.
The baryon density was sampled on {$0.3\leq\frac{n}{n0}\leq0.6$ with 8000 (or 2000, or 4000)} grid points for black (or gold, or green) curve.
The proton fraction was sampled on $0\leq x\leq0.5$ with 100 grid points.
The nuclear pressure was sampled on {$0.4\leq P_{\rm nuc}\leq0.6$} with 1000 points.
Increasing the resolution of the baryon density range is seen to {reduce the numerical fluctuations}.

Figure 7 shows the function $P_{*}(j)$ with $j$ defining the parameters according to Eq. (\ref{points100}).
Fluctuations seen on the curve are caused by the numerical error, which appears because of a finite resolution of the numerical grid that resolves the baryon density range, the proton fraction and the nuclear pressure.
The value of $P_{*}\sim 0.5\;{\rm MeV\,fm^{-3}}$ in Fig. 7 is to be contrasted with figure 2 of \cite{BBP1971}, which shows coexistence for $P_{\rm nuc}$ up to at least $1.1\;{\rm MeV\,fm^{-3}}$.

\begin{figure}
\includegraphics[width=3.5in]{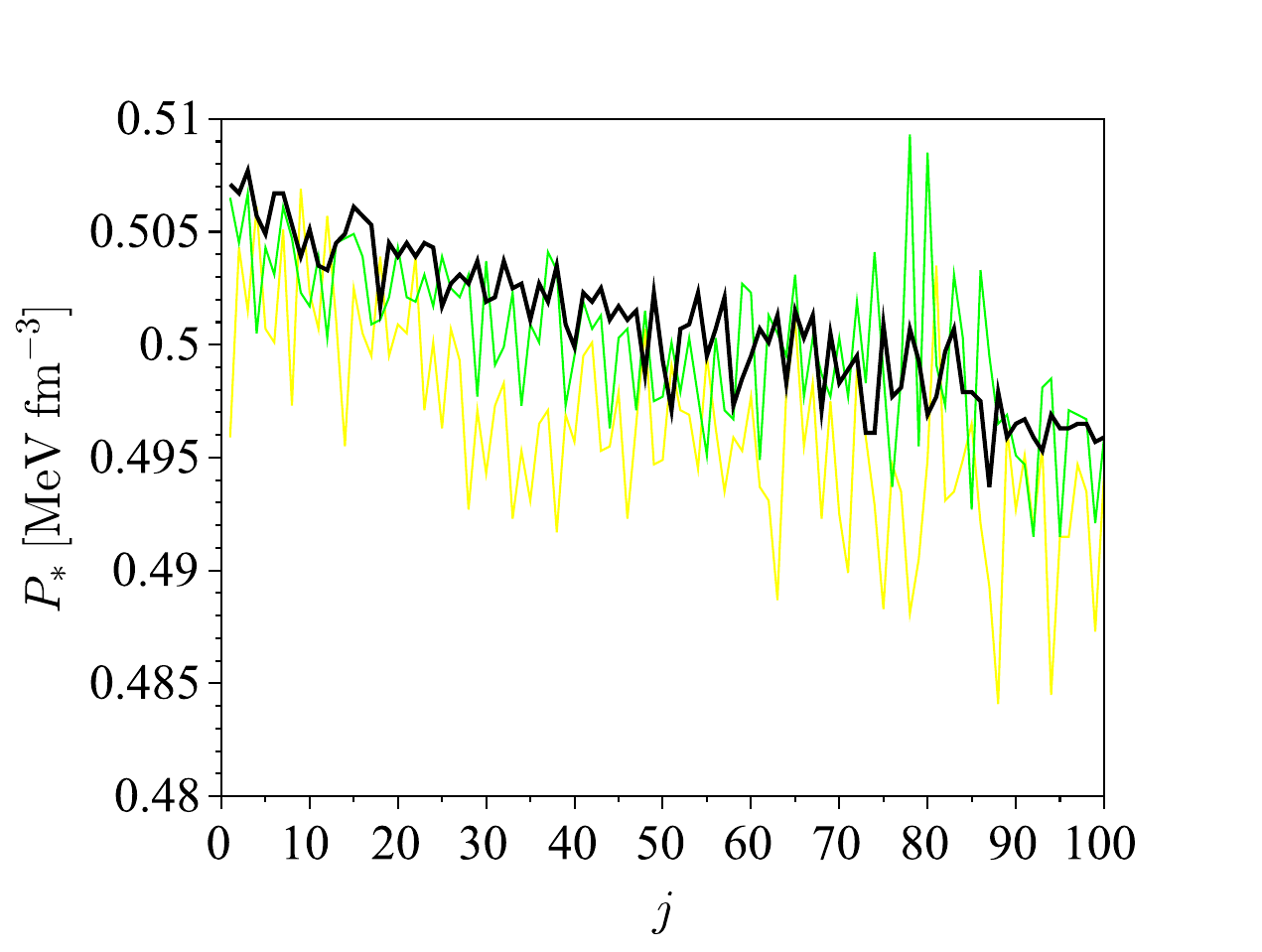}
\caption{The threshold nuclear pressure $P_{*}$ obtained from the construction used in Fig. 6, as function of $j$ defined in Eq. (\ref{points100}). Changing $j$ between 1 and 100 linearly interpolates the set of parameters $(\gamma,\alpha_L,\eta_L)$ between Eqs. (\ref{set1}) and (\ref{set2}). The coexistence is possible for $P_{\rm nuc}<P_{*}(j)$ and impossible for $P_{\rm nuc}\geq P_{*}(j)$. Black line shows calculation with 8000 grid points and green (or gold) line corresponds to 4000 (or 2000) grid points, see the text for details.}
\end{figure}

Figure 6 enables to find the difference $\Delta\mu_p$ of chemical potentials of protons in the pure neutron phase $\mu_{po}$ and in the saturated nuclear matter $\mu_{pi}$ as function of nuclear pressure:
\begin{equation}\label{Deltamup}
  \Delta\mu_p=\mu_{po}-\mu_{pi}.
\end{equation}
In the planar (slab) region of the pasta phases this quantity $\Delta\mu_p$ corresponds to a potential energy barrier for the proton to tunnel from one slab to the neighboring one.
The tunneling amplitude gives a direct measure for the anisotropic superconducting density tensor in the ordered slab pasta phase.
A rough estimate for $\Delta\mu_p\sim6$ MeV using figure 2 of \cite{BBP1971} was done by Zhang and Pethick in \cite{ZhangPethick2021}.
We now turn to detailed calculation of $\Delta\mu_p$ for various pressures.

\begin{figure}
\includegraphics[width=3.5in]{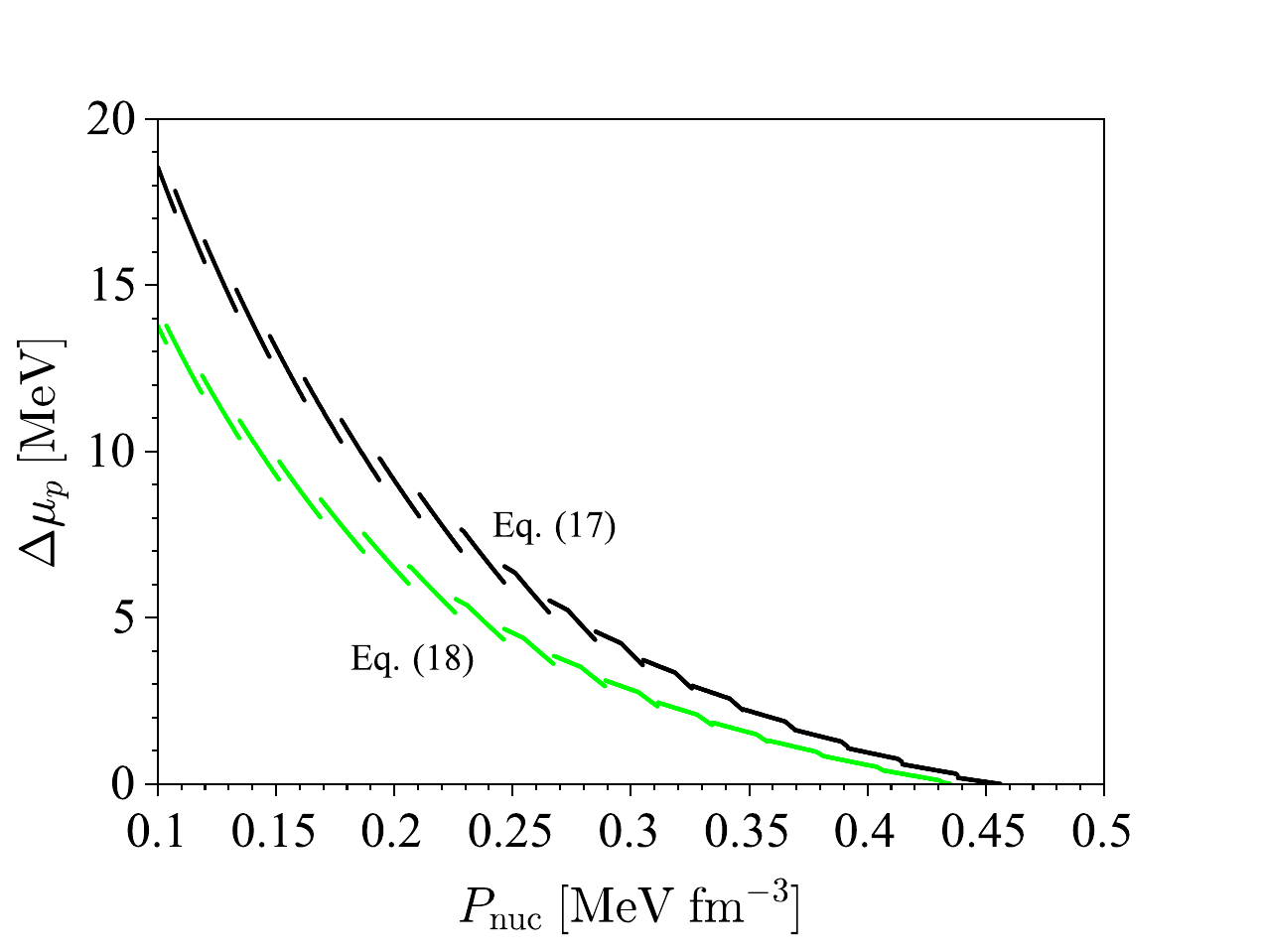}
\caption{Difference $\Delta\mu_p$ of chemical potentials of protons in the pure neutron phase $\mu_{po}$ and in the saturated nuclear matter $\mu_{pi}$ obtained from the construction of Fig. 6 as function of partial pressure $P_{\rm nuc}$ calculated using the ChEFT EoS, Eq. (\ref{e_x_n}) with the parameters $(\gamma,\alpha_L,\eta_L)$ from Eqs. (\ref{set1}) (black line) and (\ref{set2}) (green line). Step-like fluctuations of the data are caused by the numerical error.}
\end{figure}
Figure 8 shows $\Delta\mu_p$ as function of pressure $P_{\rm nuc}$ calculated with Eq. (\ref{set1}) (black line) or Eq. (\ref{set2}) (green line).
The range of nuclear pressure is chosen so that $P_{\rm nuc}<P_{*}(\gamma)$.
Fluctuations in the data are due to limited numerical accuracy.

The slab region of the pasta phase might lie in the range of baryon densities between 0.09 and 0.12 ${\rm fm^{-3}}$, as suggest Watanabe, Iida and Sato in \cite{WatanabeEtAl2000}.
In contrast, with the EoS based on ChEFT, Eq. (\ref{e_x_n}), the maximum baryon density when the coexistence (and hence, the pasta phases) is still possible, is roughly $0.5n_0=0.08\;{\rm fm^{-3}}$, which is notably below the density range found for the pasta phase by Watanabe, Iida and Sato in \cite{WatanabeEtAl2000} based on the earlier parametrization of the nuclear energy $W(k,x)$ suggested by Baym, Bethe and Pethick in \cite{BBP1971}.
It should be noted that the density range where the slab region of the pasta phase appears is quite uncertain because the numbers are very sensitive to particular model of nuclear EoS, as can be seen for instance from recent works where several EoS were compared \cite{ParmarEtAl2023,Chamel2023}.

\subsection{Superconducting density tensor at the crust-core boundary}
In the pasta phases, the superconducting density is a tensor with principal values corresponding to supercurrents along the symmetry directions of the pasta structure.
For the slab region in the ordered state and without topological defects (bridges between the adjacent sheets), the tensor has two different components: for supercurrents parallel to the normal to the sheets (or along the sheets), which is usually called $n_{\rm pp}^{\rm s \parallel}$ (or $n_{\rm p}^{\rm s \perp}$).
The difference between $n_{\rm pp}^{\rm s \parallel}$ and $n_{\rm pp}^{\rm s \perp}$ stems from the fact that supercurrents parallel to the normal to the sheets are impeded by the layers of pure neutron matter between the nuclear matter sheets.
Thus, in the slab region the superconductivity is similar to that of uniaxial superconductors observed in terrestrial layered materials.

Zhang and Pethick in equation (11) of \cite{ZhangPethick2021} evaluated the tunneling amplitude of protons through the neutron layer using a simple model of rectangular barrier for the potential energy of protons.
The barrier is caused by the difference between the proton chemical potential in the nuclear matter inside the sheet and the proton chemical potential in the pure neutron matter matter; this quantity was calculated above and presented in Fig. 8 for two parameterizations of the nuclear EoS.
Zhang and Pethick found \cite{ZhangPethick2021} that the tunneling amplitude is less than $10^{-7}$ at $P_{\rm nuc}\simeq1.1\;{\rm MeV\,fm^{-3}}$ and concluded that the tunneling is negligible.

In order to evaluate the tunneling amplitude $|T|$ I use more detailed calculations.
One of the assumptions in \cite{ZhangPethick2021} was that the lattice spacing of the slabs was approximately 44 fm and the thickness of the pure neutron layer is about 20 fm.
I use the more recent results by Watanabe, Iida and Sato \cite{WatanabeEtAl2000}, who predict in their figure 5 (with typical parameter $C_2=1$) that the thickness of the pure neutron layer is about $2(r_c-r_N)\simeq10$ fm.
Using this estimate I calculate $|T|$ from the standard formula
\begin{equation}\label{Tslabs}
  |T|\simeq4\frac{\sqrt{\Delta\mu_p\epsilon_{pi}}}{\Delta\mu_p + \epsilon_{pi}}\exp\left[-2(r_c-r)\frac{\sqrt{2m_p\Delta\mu_p}}{\hbar}\right],
\end{equation}
where $\epsilon_{pi}=\hbar^2(3\pi^2n_{pi})^{2/3}/2m_p$.
The fraction $u$ of space filled with nuclear matter is given for the slab region by $u=r/r_c$.
From figure 5 of \cite{WatanabeEtAl2000} (with typical parameter $C_2=1$) one finds $u\simeq0.5$.
\begin{figure}
\includegraphics[width=3.5in]{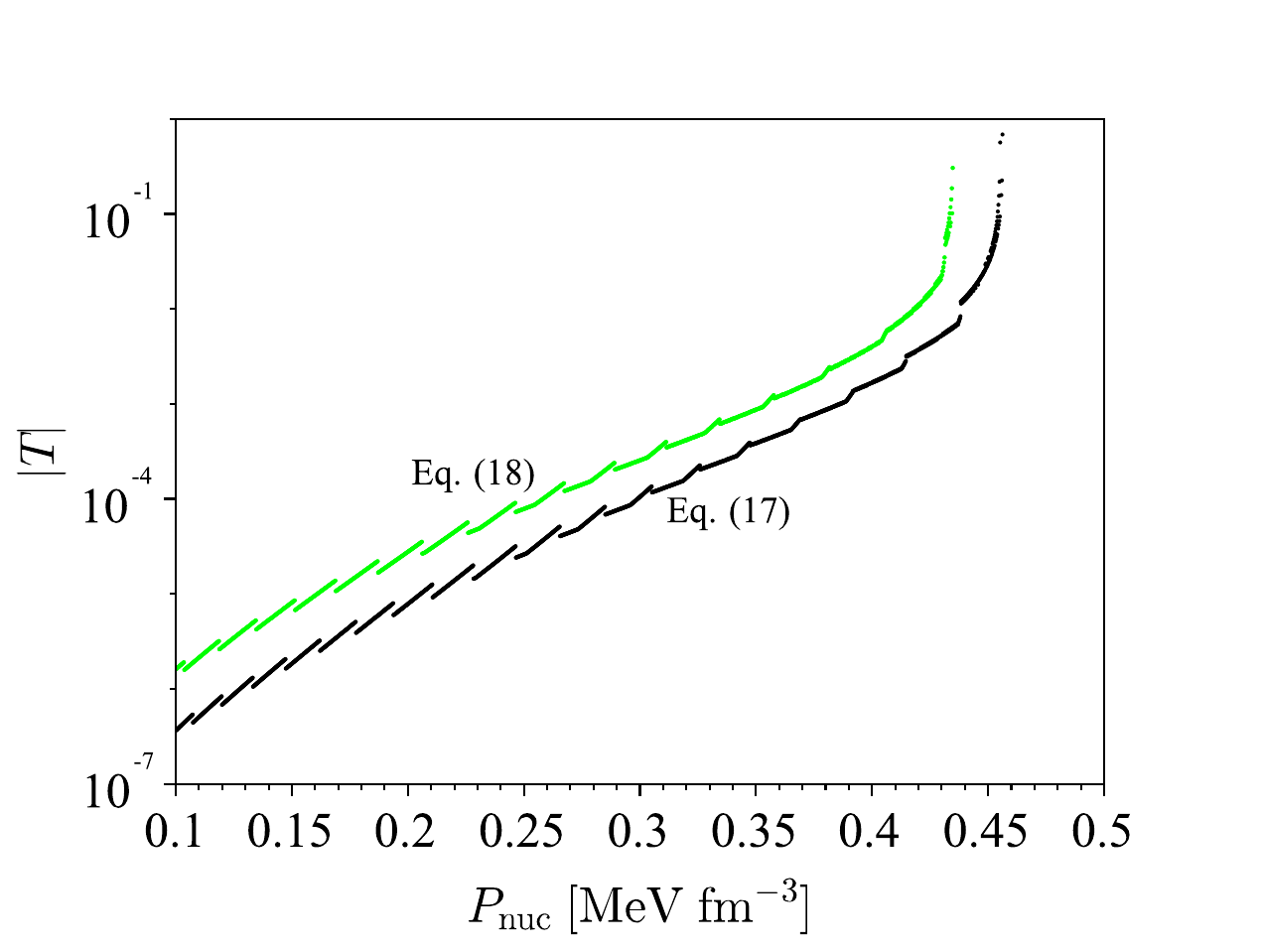}
\caption{The tunneling amplitude, Eq. (\ref{Tslabs}), of a proton between the adjacent slabs with the parameters $(\gamma,\alpha_L,\eta_L)$ from Eq. (\ref{set1}) (black line) and Eq. (\ref{set2}) (green line) in Eq. (\ref{e_x_n}). The width of the layer of the pure neutron matter between the slabs is assumed 10 fm throughout the entire slab region.
{The smallness of the tunneling amplitude implies that the flow of protons between different slabs of nuclear matter is negligible provided the structure is perfectly ordered and no bridge between the slabs is present.}
The step-like fluctuations are caused by the numerical error.}
\end{figure}

Figure 9 shows the tunneling amplitude of a proton between the adjacent slabs calculated with $\Delta\mu_p$ found in Fig. 8.
I find that $|T|$ spans a few orders of magnitude and is typically larger than the estimate obtained in \cite{ZhangPethick2021}.
However, $|T|$ remains small in most of the coexistence region and thus, the assumption of Zhang and Pethick (that the flow of protons between sheets is negligible when there are no bridges connecting adjacent sheets) is confirmed in the present calculations.
{The smallness of the tunneling amplitude implies that the flow of protons between different slabs of nuclear matter is negligible.
This property remains in power when the protons are superconducting.
In this case, the tunneling amplitude is directly related to the Josephson coupling strength between the adjacent slabs containing the superconducting protons.}

{In this paper I focus specifically on the slab region of the pasta phases, however other configurations are possible, which include the rod-like nuclei and the inverted configurations such as spherical or cylindrical bubbles.
The superconductivity is also expected in those configurations.
At present the theoretical picture of the composition of the crust-core transition is rather uncertain because the pasta phases are strongly model dependent \cite{ParmarEtAl2023,Chamel2023}.
Moreover, the thermal fluctuations become increasingly important in lower-dimensional structures where the fluctuations may significantly disturb the ordering of the structure.
This effect has been studied by Watanabe, Iida and Sato \cite{WatanabeEtAl2000} using the analogy with the liquid crystals and it was found that the structure is very sensitive to the surface energy, namely the parameter $C_2$, see Eq. (\ref{def_wsurf}).
}

{In the model used by Watanabe, Iida and Sato \cite{WatanabeEtAl2000} no magnetic torque was considered, however the magnetic field might be a stabilizing factor for the thermal fluctuations.
The magnetic torque in superconducting layered structure was considered by Zhang and Pethick \cite{ZhangPethick2021}, where they used the continuous model of anisotropic superconductivity.
Notably, Zhang and Pethick in \cite{ZhangPethick2021} assumed that the symmetry of superconductor is anisotropic and continuous.
The latter assumption is justified if the separation between the superconducting slabs is smaller than the coherence length.

Calculations of the separation distance between the adjacent slabs have been done by Watanabe, Iida and Sato in \cite{WatanabeEtAl2000}, where they used the older nuclear EoS due to Baym, Bethe and Pethick \cite{BBP1971}, which has been updated later in \cite{HebelerEtAl2013}.
Hence, the calculations of the separation distance between the adjacent slabs must be also updated.
This will be done elsewhere.

Still, the existing estimates show that the separation distance between the adjacent slabs (of the order of 10 fm) is \emph{larger} than the size of the vortex core, or the coherence length, in the superconducting order parameter (of the order of 7 fm).
In this case it is natural to describe the system as an array of discrete superconducting layers coupled through Josephson tunneling; this type of coupling is universal and is known also in Bose superfluids, which can be realized in ultacold gas \cite{Vinas2019}.
The crossover between the smooth averaged three-dimensional model (the anisotropic Ginzburg-Landau model) and the discrete two-dimensional model (the Lawrence-Doniah model), was first studied by Klemm, Luther and Beasley \cite{KlemmEtAl1975} in case of two-dimensional layers and then by Deutscher and Entin-Wohlman \cite{DeutscherEntin1978} in case of quasi two-dimensional layers.

Thus, a part of the future strategy in studies of the pasta phases is to calculate the magnetic torque in superconducting layered structure within the discreet model in order to include this factor into the model of thermal fluctuations of the pasta structure.
}

\subsection{Stellar structure}
I turn to characterization of the stellar structure using the EoS presented above.
Figure 10 shows the mass-radius relation for the three representative EoS and for the BCPM EoS obtained by solving the TOV Eqs. (\ref{TOVeq1}) and (\ref{TOVeq2}).
The EoS in the core is based either on ChEFT and extrapolated by the polytropic extension, or on the data from the BCPM EoS.
The EoS in the crust is based on the data from the BCPM EoS.
\begin{figure}
\includegraphics[width=3.5in]{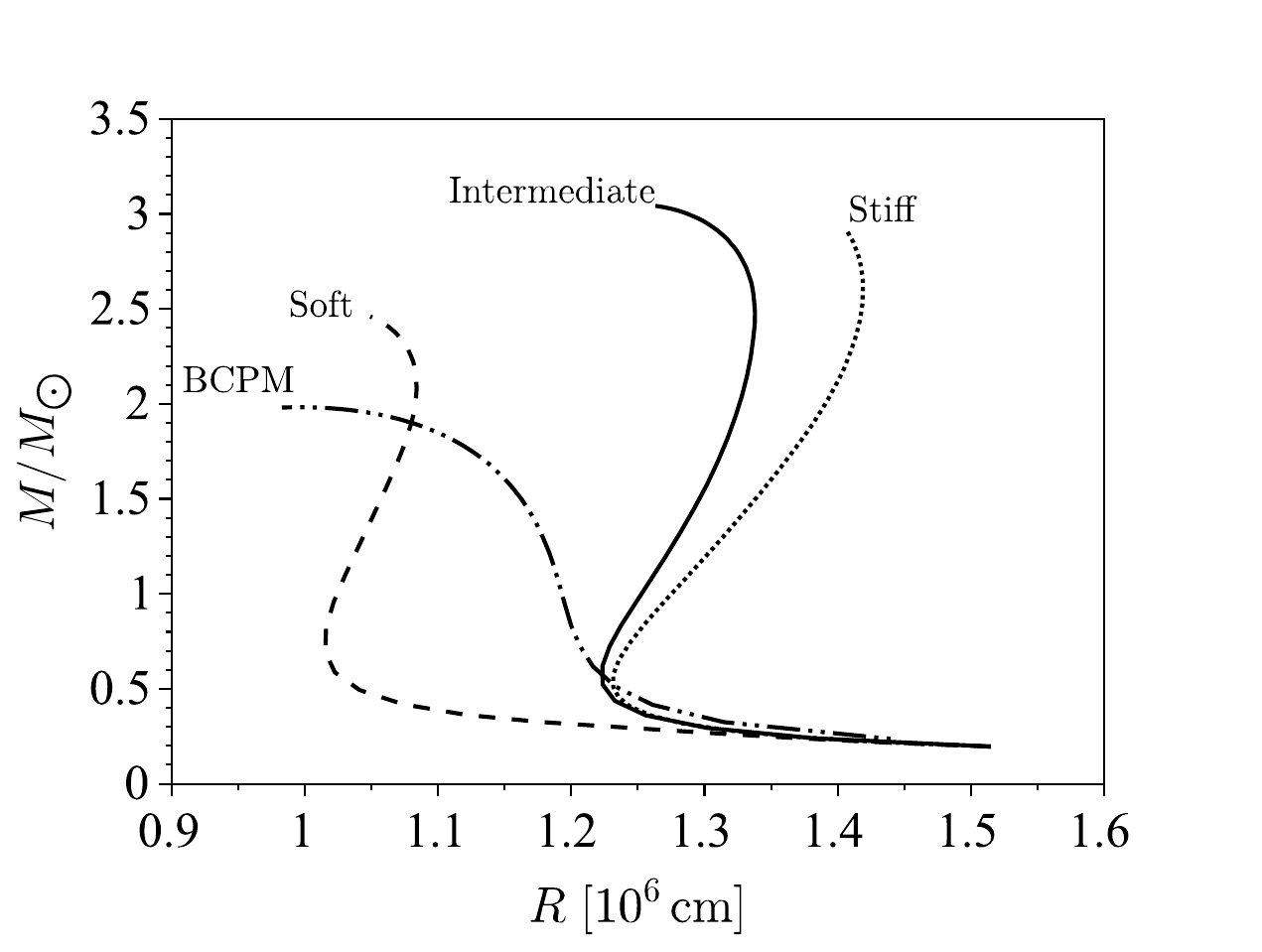}
\caption{The stellar radius $R$ in units of 10 km calculated from Eqs. (\ref{TOVeq1}) and (\ref{TOVeq2}) as function of the total stellar mass $M$ in units of the solar mass $M_{\bigodot}$. The curves correspond to different EoS in the stellar core discussed in Figs. 2 and 3 with Eq. (\ref{set1}. The crust is described by the EoS discussed in Fig. 4.}
\end{figure}

The M-R curves for representative soft, intermediate and stiff EoS correspond to the cases from the most compact to the least compact configuration.
The M-R curve generated by the BCPM EoS is located between the soft and the intermediate characteristic curves.
Comparing the M-R curve in Fig. 10 with the M-R curve obtained by Lim and Holt \cite{LimHolt2017} from the effective Skyrme interactions constrained by the ChEFT and displayed in their figure 3, it is seen that the latter corresponds to an EoS somewhere between the soft and the intermediate representative EoS shown in our Fig. 10.

One observes that the M-R relation has a notable uncertainty even for standard EoS available in the literature.
The largest uncertainty is bound to the inner core of neutron stars, where even the composition of matter is unclear.
The composition of the outer core casts much less doubts; it is expected that the outer core is composed of neutrons, protons and the electrons.

\subsection{Location of superconductivity}
The proton superconductivity is expected in the outer core \cite{BPP1969} including the crust-core boundary \cite{Kobyakov2018,KobyakovPethick2018}.
The pairing gap energy $\Delta_p$ as function of the proton Fermi wavenumber $k_{Fp}$ has been evaluated recently by Lim and Holt \cite{LimHolt2021}.
Figure 7 in \cite{LimHolt2021} shows $\Delta_p(k_{Fp})$ calculated in beta-equilibrium based on the ChEFT of nucleon interactions.
In order to establish the spatial profile $\Delta_p(r)$ in neutron star, the results for $\Delta_p(k_{Fp})$ should be translated using the spatial profile of the proton Fermi wavenumber.
The latter, $k_{Fp}(r)$, is obtained from the solution of the TOV Eqs. (\ref{TOVeq1}) and (\ref{TOVeq2}) together with the solution of the beta-equilibrium conditions which determine the composition of the matter, Eq. (\ref{betaEq}).

Combining $\Delta_p(k_{Fp})$ and $k_{Fp}(r)$ based on either ChEFT EoS or on BCPM EoS, I find the resulting $\Delta_p(r)$.
The result is shown in Fig. 11 for ChEFT EoS and the BCPM EoS.
The matter energy density scaled by $\rho_0$ and $\Delta_p(r)$ scaled by 1 MeV are shown on the same plot.
\begin{figure}
\includegraphics[width=3.5in]{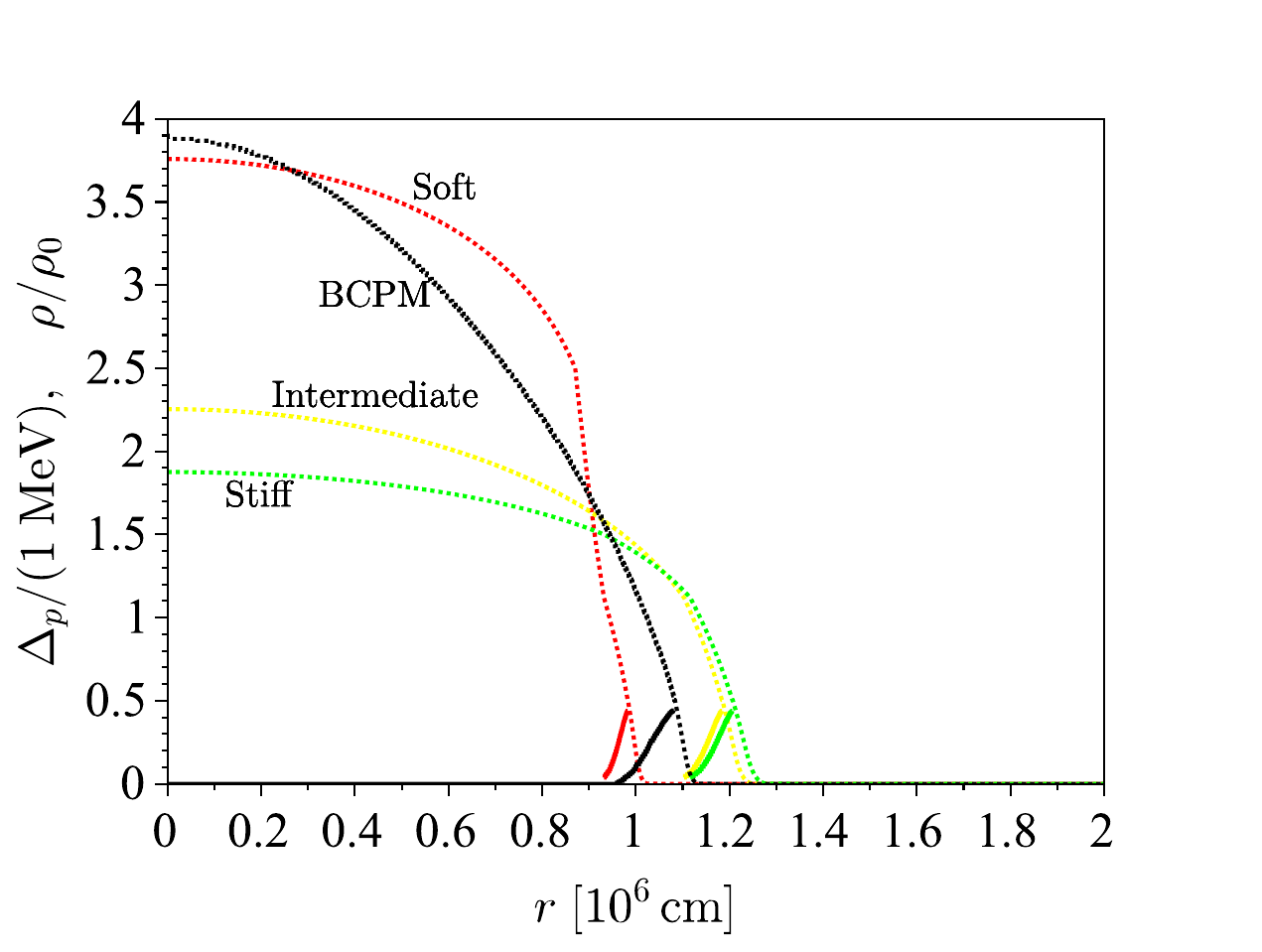}
\caption{Spatial profiles of the stellar energy density $\rho(r)$, dotted lines calculated from Eqs. (\ref{TOVeq1}) and (\ref{TOVeq2}), and of the corresponding superconducting proton pairing gap energy $\Delta_p(r)$, solid lines shown by the same color obtained by combining the solution for $\rho(r)$ and the result by Lim and Holt \cite{LimHolt2017} from their figure 7. $\rho(r)$ is calculated for three representative ChEFT-based EoS and the BCPM EoS, for the total stellar mass equal to $1.4M_{\bigodot}$. The pairing gap energy is bounded at high density by the dependence $\Delta_p(k_{Fp})$ found in figure 7 in \cite{LimHolt2021} and at low density by the crust-core boundary located at the baryon number density $n\simeq n_0/2$.}
\end{figure}
It is seen that the proton superconductivity resides in a tiny fraction of the stellar volume for any representative type of EoS.
This is the consequence of closing of $\Delta_p$ for proton Fermi wavenumbers higher than about 0.6 ${\rm fm^{-1}}$.
Thus, the superconducting region is expected to be limited by the outer core, while the inner core should be a normal electrical conductor.

\section{Magnetism of superconducting matter}
The magnetic properties of neutron star matter are strongly correlated with the location of the superconductivity and its nature.
As has been discussed above, the stellar region containing the superconducting matter is likely much narrower than usually assumed in the literature \cite{Lander2013,HenrikssonWasserman2014,MuslimovTsygan1985,GraberEtAl2015,PassamontiEtAl2017,SinhaSedrakian2015,HaskellEtAl2013,Levin2006,Bretz2021}.

The magnetic properties of the superconducting neutron star matter might be peculiar if the pasta phase is ordered.
Figure 9 suggests that in the low-density regions of the pasta phase, the tunneling amplitude between the superconducting layers is very small.
Also, since the interlayer spacing is larger than the size of the normal core of a quantized tubes of the magnetic flux, the magnetic flux can penetrate easily between the layers without a notable depletion of $\Delta_p$.
Consequently, the penetration depth of the magnetic field parallel to the slab surface{, as shown in Fig. 12 (a)} may acquire very large, macroscopic values and is limited only by the thickness of the pasta region in the stellar radial direction.

If, following \cite{LimHolt2017}, the pasta phase exists in the range of baryon number densities between 0.064 and 0.088 ${\rm fm}^{-3}$, I obtain from our Fig. 11 that the stellar layer containing the pasta phases $L_p$ has thickness of about $L_p\simeq90$ m for the soft EoS, $L_p\simeq150$ m for the intermediate EoS, $L_p\simeq160$ m for the stiff EoS and $L_p\simeq120$ m for the BCPM EoS.
Within this layer, there is a sublayer with thickness $L_s$, which contains the slab region.

Figure 12 shows the schematic setup for the perfectly ordered slab region when the magnetic field is perpendicular to the normal in panel (a) or parallel to the normal in panel (b).
The configurations shown in Fig. 12 (a) and (b) are equally possible.

If the configuration shown in Fig. 12 (a) is realized, then the magnetic field penetrates into the \emph{entire} slab region and the effective penetration depth $\lambda_\perp$ equals $L_s\sim L_p$, which is the upper bound for the penetration depth.
If the configuration shown in Fig. 12 (b) is realized, then the penetration depth is $\lambda_\parallel=\lambda_i/\sqrt{u}$, where $\lambda_i^2=m_pc^2/4\pi e^2 n_{p}$ is the London penetration depth and $e$ is the proton charge {(here, $n_{p}$ is understood as the proton density inside the dense phase of nuclear matter).
The length $\lambda_\parallel$ is the lower bound for the penetration depth.
Assuming $u=0.5$, it is easy to evaluate that $\lambda_\perp\sim L_s$ with $L_s$ between roughly $\sim 10^3$ and $\sim 10^4$ cm and $\lambda_\parallel\sim\lambda_i$ with $\lambda_i\sim10^{-11}$ cm.
This very large uncertainty uncertainty is caused by the uncertainty in mutual orientation of the magnetic field and the ordered structure, and cannot be constrained at present.

\begin{figure}
\includegraphics[width=3.5in]{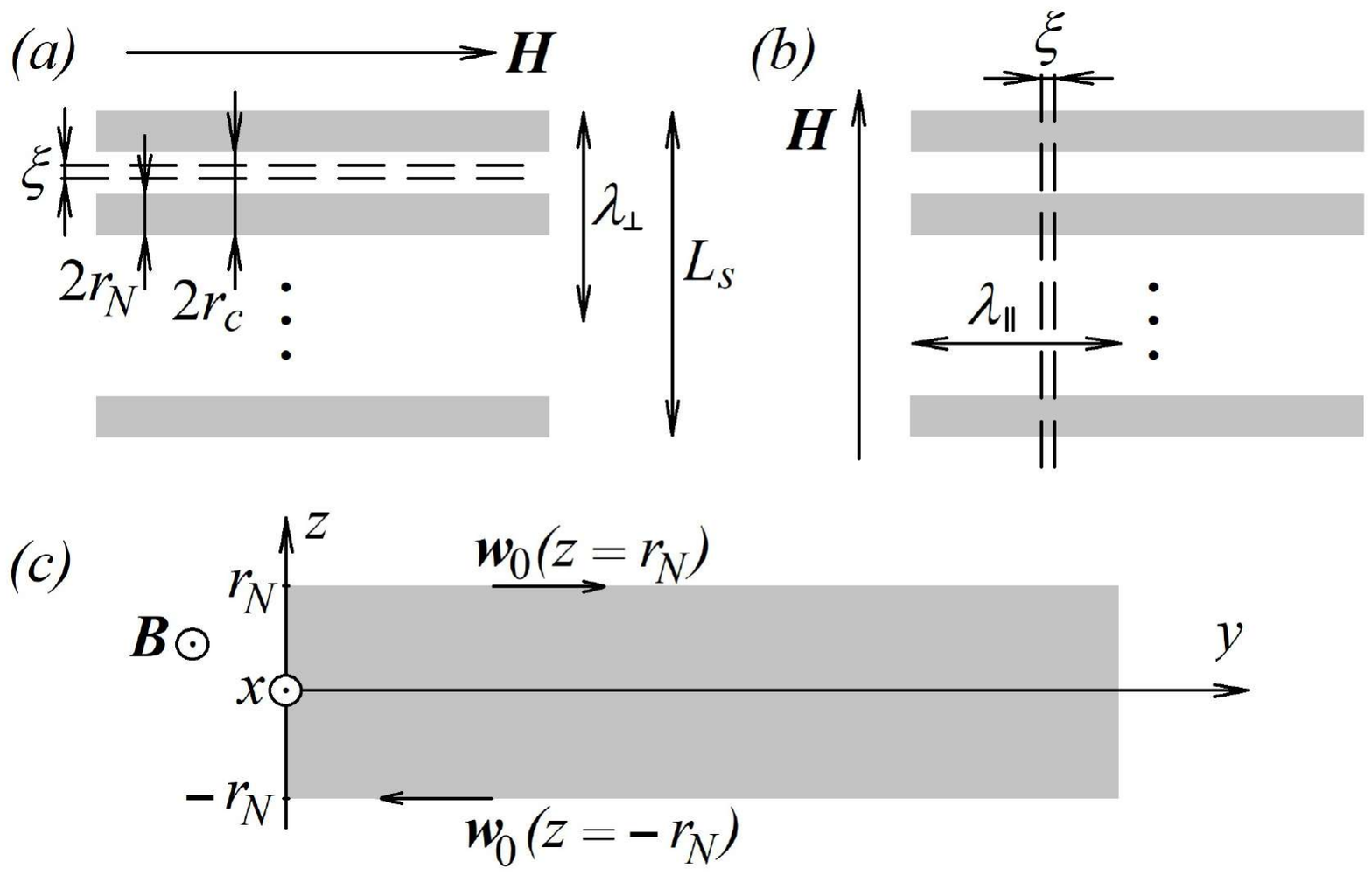}
\caption{{The schematic setup of the perfectly ordered slab region in parallel (a) and perpendicular (b) magnetic field $\mathbf{H}$.
The vortex core between the slabs (shown as grey ribbons) is depicted by dashed lines in order to emphasize that its size $\xi$ is smaller than the separation distance between the slabs.
The three black dots imply that there is a lattice of slabs with depth $L_s$.
However the vortex structure is physical only in case when it passes through the slab (grey ribbon).
Panel (c) shows a single slab in parallel magnetic field and the axes.
The unperturbed momentum lag $\mathbf{w}_0$ in stationary state acquires at the edges of the slab opposite values with equal magnitude, so that the stationary pressure drop is zero across the slab.
The force appears when a perturbation $\delta\mathbf{w}$ is superimposed on $\mathbf{w}_0$.
}}
\end{figure}

Panel (c) of Fig. 12 is a reminder of the geometry used for calculations in \cite{Kobyakov2018}.
Kobyakov \cite{Kobyakov2018} evaluated a force acting on a single slab in the magnetic field $\mathbf{B}_0=B_0\mathbf{\hat{x}}$ with assumption that the London penetration depth $\lambda_0$ corresponding to uniform saturated nuclear matter is $\lambda_0\gg r_N$.
The slabs were considered independently due to the assumption
\begin{equation}\label{SlabSeparation}
(r_c-r_N)>\xi,
\end{equation}
where $\xi$ is the coherence length of the superconductor which sets the size of the vortex core where the order parameter is depleted, with $\xi\ll\lambda_0$.

{The force found by Kobyakov in \cite{Kobyakov2018} can be understood as an additional pressure due to nonzero velocity lag between the neutrons and protons.
As a reminder I briefly review the calculations in \cite{Kobyakov2018}.
The spatial density of force acting on a single slab is given by equation (29) in \cite{Kobyakov2018}.
The total force can be found by integration of the momentum flux tensor in the superconducting-superfluid mixture
\begin{eqnarray}
\nonumber &&  \Pi_{ik}=J_{pk}\pi_{pi} + J_{nk}p_{ni} + \delta_{ik}\left(p - \rho_{np}^{*}\frac{\mathbf{w}^2}{2}\right) \\
 \label{MFtensor} &&  + \frac{1}{4\pi}\left(\delta_{ik}\frac{\mathbf{B}^2}{2}-B_iB_k\right).
\end{eqnarray}

Here, $ik$ are the Cartesian indices, $\mathbf{J}_{p}$ and $\mathbf{J}_{n}$ (or $\mathbf{p}_{p}$ and $\mathbf{p}_{n}$) are the number currents (or superfluid momenta equal to gradient of the superfluid order parameter phase) of protons and neutrons correspondingly, $p$ is the pressure without the contributions due to the matter flows as given by equation (32) in \cite{Kobyakov2018}, $\pmb{\pi}_{p}=\mathbf{p}_{p}-e\mathbf{A}/c$ is the gauge-invariant proton momentum, $\mathbf{A}$ is the electromagnetic vector potential, $\mathbf{w}=(\pmb{\pi}_{p}-\mathbf{p}_{n})/m$ is the momentum lag, with $m=m_p\approx m_n$, $\rho_{np}^{*}$ is the entrainment mass density and $\mathbf{B}=\nabla\times\mathbf{A}$.
Using the geometry shown in Fig. 12 (c), the vector potential is $\mathbf{A}_0=-\mathbf{\hat{y}}B_0z$ and the unperturbed superfluid gauge-invariant momentum lag is given by equation (35) in \cite{Kobyakov2018}, or $\mathbf{w}_0=(e/mc)\mathbf{\hat{y}}B_0z$, it is also displayed in Fig. 12 (c).

It is clear that the dynamical degrees of freedom are spanned by the variables $(\mathbf{v}_p,\mathbf{w},\mathbf{B})$, where $\mathbf{v}_p=\mathbf{J}_{p}/n_p$.
Therefore each of these variables can be perturbed independently.

I assume that $\mathbf{w}$ is perturbed according to $\mathbf{w}=\mathbf{w}_0+\delta\mathbf{w}$, where $\delta\mathbf{w}=\delta w(\mathbf{\hat{y}}\cos\theta+\mathbf{\hat{x}}\sin\theta)$.
This form of perturbation generalizes equation (38) in \cite{Kobyakov2018} to arbitrary direction of $\delta\mathbf{w}$ relative to $\mathbf{w}_0$ for the configuration when the magnetic field is perpendicular to the normal of the slab.
Noting that the relevant force is given by
\begin{equation}\label{Ftot1}
\delta^{(1)}\mathbf{F}_{\rm tot}^{1\,{\rm cm}^2\times2r_c}=\mathbf{\hat{z}}\int d^3\mathbf{r} \sum_k\nabla_k\Pi_{3k},
\end{equation}
and using the hydrodynamic pressure $P=p-\rho_{np}^{*}\mathbf{w}^2/2$, I find
\begin{eqnarray}
\nonumber && \delta^{(1)}\mathbf{F}_{\rm tot}^{1\,{\rm cm}^2\times2r_c}=1\,{\rm cm}^2\times\left[P(z=r) - P(z=-r)\right] \\
\label{Ftot2} && =-1\,{\rm cm}^2\times\mathbf{\hat{z}}\frac{e}{c}B_02rn_{np0}\delta w\cos\theta.
\end{eqnarray}
The hydrostatic pressure $p$ is taken constant across the slabs in order to calculate the additional pressure due to the peculiar force.
The crustal stress arises as soon as we impose condition that the slab is static due to external forces at the boundary of the domain; in this case $p$ adjusts in order to compensate the additional pressure, and this affects the resulting momentum flux at the domain boundary, which physically implies the crustal stress.
Equation (\ref{Ftot2}) generalizes the result obtained for the first time by Kobyakov in \cite{Kobyakov2018}.
Note that the factor of $(n_{n0}-n_{n0}^o)/n_{n0}$ in equation (40) in \cite{Kobyakov2018} is erroneous, but this correction does not affect the conclusions.
}

\section{Conclusions}
Based on calculations presented in this paper the following conclusions can be drawn.
The superconducting matter in neutron stars calculated from reliable modern knowledge about the nuclear matter fills only a tiny fraction of the star and is located somewhere at the tip of the saturated nuclear matter in a layer of thickness about 600-650 meters, as shows Fig. 11 and is explained in Sec. IV.
From the bottom of the layer, about 500 meters are filled by the isotropic superconductor, and the rest 100-150 meters might be filled with a superconducting liquid crystal.

If the state of the liquid crystal is ordered, the superconductor has a discreet lattice symmetry with one axis in the slabs region, or two axes in the rod-like nucleus region.
The discreet nature of the symmetry follows from the results of the calculations with the chosen EoS, which have shown that the tunneling rate of protons between the slabs is negligible and that the superconductor coherence length is smaller than the separation between the adjacent slabs, as shown in Fig. 9 and explained in Sec. III B.
However, the geometry of the pasta phases is very sensitive to the chosen model of the nuclear interactions and should be theoretically constrained as soon as a better coherence between the various models of nuclear forces is achieved.

The existing calculations have shown that the pasta structure is expected in the range of densities roughly between 0.08 and 0.12 ${\rm fm}^{-3}$ as shows figure 5 of \cite{WatanabeEtAl2000}, or between 0.07 and 0.08 ${\rm fm}^{-3}$ as it can be seen in table 5 of \cite{SharmaEtAl2015}, or between 0.06 and 0.09 as seen in figure 6 of \cite{LimHolt2017}.
In this paper I have shown in Fig. 7 that with the ChEFT EoS, the coexistence is possible for pressure up to roughly 0.5 ${\rm MeV\,fm}^{-3}$.
This implies, as shows Fig. 3, that the maximum baryon density at which the pasta phase is possible, is roughly $0.59n_0=0.0944$ ${\rm fm}^{-3}$ for the parametrization from Eq. (\ref{set1}) and roughly $0.54n_0=0.0864$ ${\rm fm}^{-3}$ for the parametrization from Eq. (\ref{set2}).
These values are consistent with the calculations in the framework of Skyrme interactions constrained by ChEFT \cite{LimHolt2017}, as shows their figure 6, and are consistent with the calculations based on BCPM \cite{SharmaEtAl2015} as shows their table 5.
The coexistence calculations exclude the explicit calculations of the surface and Coulomb corrections which have been shown to be small in Sec. II D and thus, represent a simple and reliable cross-check to constrain the crust-core transition structure.

Magnetism of neutron stars is strongly correlated with the superconductivity and is considerably uncertain due to uncertainties related to the pasta structure.
One of the most obscured property is the ordering of the pasta phase.
A strategy to constrain the ordering consists of parallel efforts in calculation of the geometrical parameters of the pasta and in studies of the thermal fluctuations.
In ordered lower-dimensional structures, thermal fluctuations can be studied in analogy with liquid crystalline matter.
However, interaction with the magnetic field must be included, which requires to calculate the torque on the superconducting pasta structure induced by the magnetic field.
Initial estimates performed in this paper, including the calcuations of the chemical potentials in the context of coexistence and the corresponding proton tunneling rate between the slabs, suggest that the discreet model of superconductivity rather than a continuous one should be used to calculate energy of the magnetic flux tube in superconducting pasta.
This result validates from microscopic physics the assumption of independent slabs in the ordered pasta and leads to a possibility of macroscopically large penetration depth of the magnetic field in the configuration shown in Fig. 12 (a) as explained in Sec. IV.

Another significant uncertainty in the stellar magnetism arises in the context of the magnetoacoustic waves in the interior of neutron stars.
Profound effects of superconductivity on the magnetoacoustic waves are expected because these waves propagate with different speeds in superconducting and in normal plasma.
Calculations of this paper have shown that the core is not entirely superconducting.
Therefore somewhere in the core a normal-superconducting boundary exists.
This implies that the normal magnetized plasma below this boundary will be effectively confined in a superconducting cavity, because with the magnetic field frozen into the normal plasma, the plasma motion across the normal-superconducting boundary is suppressed due to the Meissner screening.
The earlier models of the global magnetic energy transfer in the interior may be improved by taking into account this prediction.

By bringing together aspects of knowledge about the superconducting properties in different stellar regions, in this paper I paved the way to develop a unified picture of superconductivity in neutron stars.
The unified description of superconductivity sets the stage for tracing the relations between the internal features of the magnetism, the neutron star structure and the astronomically observed signals.

\section*{Acknowledgements}
I am grateful for the support by Center of Excellence ``Center of Photonics'' funded by The Ministry of Science and Higher Education of the Russian Federation, contract № 075-15-2022-316.

\end{document}